# Role of NMDA conductance in average firing rate shifts caused by external periodic forcing

Nikita Novikov [1] and Boris Gutkin [1,2]

[1] *Centre for Cognition and Decision Making, National Research University Higher School of Economics, Moscow, Russia, 101000*
[2] *Group for Neural Theory, LNC INSERM U960, Department of Cognitive Studies, Ecole Normale Superieure PSL\* Research University, Paris, France, 75005*

## ABSTRACT

A widely accepted view of computations in the brain relies on population coding, where the neural ensemble firing rate is modulated in a stable manner to transmit information and perform various cognitive tasks. At the same time, oscillatory neural activity is specifically modulated in frequency, coherence and power during cognitive performance. How the firing rate and oscillations interact remains a salient question. In this paper, we develop a theory for the interactions between oscillatory signals and the firing rate of neural populations based on activity of non-linear voltage-dependent NMDA synapses. Notably, we show under which conditions oscillatory inputs can control the mean firing rate without loss of stability. Using mathematical analysis and simulations of mean-field models, we demonstrate that presence of NMDA synapses on both the excitatory and the inhibitory neurons is critical for sinusoidal oscillations to significantly and stably increase the firing rate. We characterize the oscillation-induced mean firing rate shift as a function of the fast and slow synaptic weights and demonstrate the parameter region, in which the effect under investigation is mostly pronounced. Results of our work may help identify the properties of neural circuits that allow for constructive control of the firing rate codes by large-scale neural oscillations.

## I. INTRODUCTION

Oscillatory activity is abundant across neural networks of the brain, and its profile is sensitive to various experimental and natural conditions (Wang, 2010). However, the relation between neural oscillations and the most basic functions of the brain circuits such as encoding and redistributing of information is poorly understood. One of the most prominent principles of information encoding in the brain is population rate coding (Rolls, 2011). According to this principle, information is stored as distribution of firing rates across a population. Firing rate in this case is defined as the average number of spikes emitted in a relatively prolonged time interval. In this context, a question of how might the firing rate modulations, necessary for information transmission and task execution, may interact with the oscillatory profile of the neuronal activity, comes to the forefront.

In many cases, oscillatory profile is closely related to population firing rate (Azouz, Gray, 2003). Also, there is evidence that oscillatory activity is not merely a byproduct of rate coding, as it could be modulated independently of firing rates (Fries et al., 2001). In principle, oscillatory activity could be related to rate coding in several ways. First, it could serve as an independent dimension of neural code. Second, it could serve as a "metadata" that controls transmission of rate-coded information in the brain (Akam, Kullmann, 2014; Fries, 2005). Third, it could provide a modulatory scaffold for rate coding which stabilizes the code or makes the coding impossible by destabilizing the necessary firing rate dynamics (Dipoppa, Gutkin, 2013; Schmidt et al., 2018).

In order to algorithmically implement the aforementioned functions, there should be mechanisms of transformation between population rate code and oscillatory profile (Akam, Kullmann, 2010; Akam, Kullmann, 2014). The conversion from firing rate to oscillatory properties is quite well understood at the level of simple spiking networks (Brunel, Wang, 2003); also certain biophysical mechanisms were described for microcircuits (Roopun et al., 2008), as well as for single neurons. However, the opposite conversion, i.e. the effect of oscillatory properties on the mean firing rate, is much less studied.

To change the mean firing rate of a system by oscillatory entrainment, the system should be non-linear. One candidate property is non-linearity of the f-I curves (gain functions) of neurons in the subthreshold state. The positive half-wave of input oscillations brings a neuron close to the threshold, thus increasing the firing probability, while the negative half-wave does not strongly affect this probability.

Thus, in average, oscillatory input modulation could increase the mean firing rate (Rolls, 2011; Salinas, Sejnowski, 2000); this effect has been explored mathematically in the limit of weak oscillations (Voronenko, Lindner, 2017). However, given that in very noisy state (typical for cortical networks) the gain functions are close to linear, the aforementioned mechanism should produce quite subtle effect.

Another possibility to link oscillations and firing rate modulation is to consider slow non-linear mechanism, such as NMDA conductance or voltage-dependent ion channels; this idea was proposed in the study (Akam, Kullmann, 2010), but has not been implemented in that study. In the present paper, we theoretically investigate the role that slow voltage-dependent NMDA currents could play in shifting the time-averaged state of an excitatory-inhibitory system in the presence of external zero-mean sinusoidal forcing. In order to separate the effect of NMDA non-linearity from the aforementioned effect of gain function non-linearity, we consider a low-dimensional population model with the gain functions linearized about the unforced equilibrium. Slopes of the gain functions, as well as the time constants that define the dynamics of population firing rates and voltages, are numerically derived from simulations of the uncoupled networks of leaky integrate-and-fire neurons. In our study, we mainly concentrate on excitatory effect produced by the external forcing. We demonstrate that in the presence of NMDA-receptors on the excitatory neurons only, the positive shift of the firing rate could only be very slight (1-2 Hz) without loss of stability, while the positive inhibitory firing rate shift could be more pronounced. However, when NMDA-receptor are also present on the inhibitory neurons, the system could be made more excitable without loss of stability, and the large positive shifts of both excitatory and inhibitory firing rates could be achieved.

The paper is organized as follows. (1) We describe our model and its linearized version. (2) We provide analytical expressions for the relation between the parameters of the external forcing and the shift of the time-averaged state of the system that the forcing produces. (3) We describe the procedure of parameter selection that provides a realistic state of the system. (4) We perform geometrical analysis of the phase plane for the system with NMDA-receptors on the excitatory population only. We find the time-averaged steady-states for the unforced and for the forced system. We also discuss the stability-related limitations for the steady-state shift that follows from our geometrical consideration. (5) We confirm predictions of our geometrical analysis by explicitly simulating the corresponding population model. (6) We perform the geometrical analysis and the confirming simulation for the system with NMDA-receptors on both populations (excitatory and inhibitory). We discuss how adding NMDA-receptors to the inhibitory population helps to overcome the limitations on the steady-state shift. (7) Finally, we compare the two aforementioned cases for a range of values of a system parameter that controls excitability of the fast subsystem.

## II. THEORY

### A. The low-dimensional neural circuit system model

In order to describe the neural population dynamics of an excitatory-inhibitory neural circuit we use a low dimensional model. The two different neural populations are recurrently intercoupled with instantaneous excitatory AMPA and inhibitory GABAA, as well as slow non-linear excitatory NMDA connections. The network also receives external excitatory inputs represented by sum of white-noise signal with non-zero mean and sinusoidal signal with zero mean. We describe our system by the following six variables:

(1) the excitatory and inhibitory population firing rates: $r_e, r_i$,

(2) the mean membrane voltages averaged over the excitatory and the inhibitory populations: $V_e, V_i$, and

(3) the mean NMDA currents received by the excitatory and the inhibitory neurons: $I_{NMDAe}, I_{NMDAi}$; these currents depend on the presynaptic firing rate, as well as on the postsynaptic population voltage.

Dynamics of these variables are given by a system of first-order differential equations with the variables evolving around their state-dependent instantaneous "equilibrium" values $r_e^{ss,0}, r_i^{ss,0}, V_e^{ss,0}, V_i^{ss,0}, I_{NMDAe}^{ss,0}, I_{NMDAi}^{ss,0}$ on time scales defined by the time constants $\tau_{re}, \tau_{ri}, \tau_{Ve}, \tau_{Vi}, \tau_{NMDA}$:

$$\begin{cases} \dfrac{dr_e}{dt} = \dfrac{1}{\tau_{re}} \left( r_e^{ss,0}(u_e(r_e,r_i,I_{NMDAe})) - r_e \right) \\ \dfrac{dr_i}{dt} = \dfrac{1}{\tau_{ri}} \left( r_i^{ss,0}(u_i(r_e,r_i,I_{NMDAi})) - r_i \right) \\ \dfrac{dV_e}{dt} = \dfrac{1}{\tau_{Ve}} \left( V_e^{ss,0}(u_e(r_e,r_i,I_{NMDAe})) - V_e \right) \\ \dfrac{dV_i}{dt} = \dfrac{1}{\tau_{Vi}} \left( V_i^{ss,0}(u_i(r_e,r_i,I_{NMDAi})) - V_i \right) \\ \dfrac{dI_{NMDAe}}{dt} = \dfrac{1}{\tau_{NMDA}} \left( I_{NMDAe}^{ss,0}(r_e,V_e) - I_{NMDAe} \right) \\ \dfrac{dI_{NMDAi}}{dt} = \dfrac{1}{\tau_{NMDA}} \left( I_{NMDAi}^{ss,0}(r_e,V_i) - I_{NMDAi} \right) \end{cases} \quad (1)$$

Here, $u_e, u_i$ denote the mean instantaneous inputs to the excitatory and the inhibitory populations respectively (expressed in terms of voltage). The functions $r_e^{ss,0}(u_e), r_i^{ss,0}(u_i), V_e^{ss,0}(u_e), V_i^{ss,0}(u_i)$ take the mean inputs as its arguments and depend both on the properties of the neurons and on the level of noise in the system (we derive these functions from simulations of individual neurons, see below). Please note that we do not model the noise explicitly; instead, its effect is incorporated in the functions $r_e^{ss,0}(u_e), r_i^{ss,0}(u_i), V_e^{ss,0}(u_e), V_i^{ss,0}(u_i)$. The functions giving the nonlinear NMDA-dependent excitation ($I_{NMDAe}^{ss,0}(r_e,V_e), I_{NMDAi}^{ss,0}(r_e,V_i)$) take the excitatory firing rate and the mean voltage of the input population as its arguments and are expressed as follows:

$$I_{NMDAa}^{ss,0}(r_e,V_a) = J_{ae}^{NMDA} g_{NMDA}(V_a) r_e, \quad (2)$$

where the index $a$ could be replaced by $e$ for the excitatory population or by $i$ for the inhibitory population (this notation will be used along the paper). Here $J_{ae}^{NMDA}$ is the strength of the NMDA-related coupling between the excitatory population and the population $a$; the function $g_{NMDA}$ describes the dependence of NMDA-current on the membrane voltage:

$$g_{NMDA}(V) = \left[1 + \exp(-0.062 \cdot V)/3.57\right]^{-1}, \quad (3)$$

where $V$ is expressed in millivolts.

The inputs $u_e, u_i$ depend on the firing rates $r_e, r_i$, the external tonic inputs $h_e, h_i$, and the NMDA-currents $I_{NMDAe}, I_{NMDAi}$ as follows:

$$u_a(r_e,r_i,I_{NMDAa}) = J_{ae}^{AMPA} r_e - J_{ai} r_i + h_a + I_{NMDAa}/g_{ma}, \quad (4)$$

where $J_{ae}^{AMPA}, J_{ai}$ are the strengths of the fast synaptic couplings (the first index defines the input population, and the second index – the output population), $g_{ma}$ is the membrane conductance of neurons from the population $a$. The synaptic coupling strengths are expressed as follows:

$$\begin{cases} J_{ae}^{AMPA} = j_{ae}^{AMPA} K_{ae} \tau_{ma} \\ J_{ae}^{NMDA} = j_{ae}^{NMDA} K_{ae} \tau_{NMDA} \\ J_{ai} = j_{ai} K_{ai} \tau_{ma} \end{cases} \quad (5)$$

where $j_{ai}$ is the amplitude of the instantaneous inhibitory postsynaptic potential on the neurons from the population $a$; $j_{ae}^{AMPA}$ is the amplitude of the instantaneous excitatory AMPA-postsynaptic potential onto the neurons in the population $a$; $j_{ae}^{NMDA}$ - the amplitude of the NMDA-postsynaptic current step onto the neurons from the population $a$; $K_{ab}$ - number of inputs that a neuron from population $a$ receives from neurons that belong to population $b$ (a,b = e,i), $\tau_{ma}$ - membrane time constant the neurons from the population $a$.

The time constant of NMDA-synaptic input is much larger than the time constants that govern the dynamics of the population firing rates and the membrane voltages, i.e. $\tau_{NMDA} \gg \tau_{re}, \tau_{ri}, \tau_{Ve}, \tau_{Vi}$. Thus, we can consider our system as a slow-fast system, and separate the time scales. At the fast time scale, the state of the slow subsystem $(I_{NMDAe}, I_{NMDAi})$ can be considered as constant. Then the fast subsystem $(r_e, r_i, V_e, V_i)$ converges to an equilibrium $(r_e^*, r_i^*, V_e^*, V_i^*)$ that depends on the state of $(I_{NMDAe}, I_{NMDAi})$. At the slow time scale, the slow subsystem $(I_{NMDAe}, I_{NMDAi})$ evolves, and its dynamics assumes that the fast system is converged to an instantaneous equilibrium at each time moment, i.e. $(r_e, r_i, V_e, V_i) = (r_e^*, r_i^*, V_e^*, V_i^*)$. As seen below, we use the time scale separation to analyze the stability of the fast and the slow subsystems independently, as well as to analyze the effect of the forced oscillations of the fast subsystem on the dynamics of the slow subsystem.

## B. Modelling assumptions

In the model described above, we made the following assumptions: (1) fast AMPA and GABAA synapses are instantaneous; (2) NMDA currents are used as dynamic variables, instead of NMDA-conductances; (3) NMDA-currents are non-saturating; (4) voltage and firing rate dynamics are described by linear differential equations, without taking the effect of spike-to-spike synchronization into account; (5) time constants $\tau_{re}, \tau_{ri}, \tau_{Ve}, \tau_{Vi}$ do not depend on the state of the system; (6) the main source of noise is external, and it is accounted for in the functions $r_e^{ss,0}(u_e), r_i^{ss,0}(u_i), V_e^{ss,0}(u_e), V_i^{ss,0}(u_i)$. The assumption (4) is valid when the system operates in a subthreshold regime, in which the activity of the corresponding spiking network is noise-driven and irregular (in this work, we check that the parameters we use satisfy these conditions). Note that this regime is different from a supra-threshold one, where the corresponding spiking network would show regular oscillatory spiking. We an argue that the assumption (6) is valid if the fast synaptic weights are sufficiently small, which is also true in our case. While assumptions (1) and (5) may be problematic in general, they only affect the relationship between the external forcing and the forced oscillations of the fast system. In our analysis, we start from selecting the parameters of the forced oscillations (as opposed to deriving them from the parameters of the external forcing), so the aforementioned relationships are not important in our case. The assumptions (2) and (3) are crucial, as they affect the non-linear dependence of the NMDA-currents on the firing rates and the membrane voltages. We use these assumption to simplify our analysis.

We should also note that we the expression (3) for $g_{NMDA}(V)$ is valid for single-neurons. In the network, there is a distribution of voltages across neurons $\{V_a^k\}$ ($a=e,i$), where $k$ is a neuron number; the mean voltage of a population a is denoted by $\langle V_a^k \rangle_k = V_a$. Therefore, each neuron is characterized by its own value of the voltage-dependent NMDA conductance $g_{NMDA}(V_a^k)$. In the expression for the mean NMDA-current across neurons, one should use the term $\langle g_{NMDA}(V_a^k) \rangle_k$, we make a simplifying assumption and use $g_{NMDA}(\langle V_a^k \rangle_k) = g_{NMDA}(V_a)$. Of course, from the non-linearity of $g_{NMDA}$, it follows that $g_{NMDA}(\langle V_a^k \rangle_k) \neq \langle g_{NMDA}(V_a^k) \rangle_k$. However, under a quasi-static approximation, it should be possible to find the appropriate function $\tilde{g}_{NMDA}$ such that $\tilde{g}_{NMDA}(V_a) = \langle g_{NMDA}(V_a^k) \rangle_k$. Yet this would make our analysis unwieldy while having presumably small effect on our main conclusions.

## C. The linearized system

The goal of our study is to investigate the impact of zero-mean periodic forcing on shift of the equilibrium. In the system (1), this shift can be accounted for by non-linearity of the functions $r_a^{ss,0}, V_a^{ss,0}$

(governing the dynamics of the fast subsystem), as well as by non-linearity of the functions $I_{NMDAa}^{ss,0}$. In this paper, we concentrate on the effect of the $I_{NMDAa}^{ss,0}$ non-linearity. Thus, to analyze this effect separately, we linearize $r_a^{ss,0}, V_a^{ss,0}$ about the equilibrium. We should note, however, that the assumption of linearity of $r_a^{ss,0}, V_a^{ss,0}$ is non-realistic, especially in the case when the induced oscillations are strong; in the realistic situation the influence of both $r_a^{ss,0}, V_a^{ss,0}$ and $I_{NMDAa}^{ss,0}$ non-linearities on the forcing-induced equilibrium shift should be considered.

We now describe the linearization procedure. Let us assume that the full system has a stable fixed point denoted as $(r_{e0}, r_{i0}, V_{e0}, V_{i0}, I_{NMDAe0}, I_{NMDAi0})$ and the corresponding values of inputs as $(u_{e0}, u_{i0})$. Now let us linearize the functions $r_e^{ss,0}(u_e), r_i^{ss,0}(u_i), V_e^{ss,0}(u_e), V_i^{ss,0}(u_i)$ about this fixed point, and substitute these functions in (1) by their linearized versions, which we denote as $r_e^{ss}(u_e), r_i^{ss}(u_i), V_e^{ss}(u_e), V_i^{ss}(u_i)$, respectively (let us redefine $I_{NMDAe}^{ss}(r_e, V_e) \equiv I_{NMDAe}^{ss,0}(r_e, V_e)$, $I_{NMDAi}^{ss} \equiv I_{NMDAi}^{ss,0}(r_e, V_i)$ for notational convenience):

$$\begin{cases} \dfrac{dr_a}{dt} = \dfrac{1}{\tau_{ra}} \left( r_a^{ss}(u_a(r_e, r_i, I_{NMDAa})) - r_a \right) \\ \dfrac{dV_a}{dt} = \dfrac{1}{\tau_{Va}} \left( V_a^{ss}(u_a(r_e, r_i, I_{NMDAa})) - V_a \right) \\ \dfrac{dI_{NMDAa}}{dt} = \dfrac{1}{\tau_{NMDA}} \left( I_{NMDAa}^{ss}(r_e, V_a) - I_{NMDAa} \right) \end{cases} \quad (6)$$

Below, we will refer to (6) as the "unforced system".
The linearized functions are expressed as follows:

$$\begin{cases} r_a^{ss}(u_a) = r_{a0} + c_{ra} \cdot (u_a - u_{a0}) \\ V_a^{ss}(u_a) = V_{a0} + c_{Va} \cdot (u_a - u_{a0}) \end{cases}, \quad (7)$$

where the appropriate derivatives are:

$$\begin{cases} c_{ra} = \left. \dfrac{dr_a^{ss,0}(u_a)}{du_a} \right|_{u_{a0}} \\ c_{Va} = \left. \dfrac{dV_a^{ss,0}(u_a)}{du_a} \right|_{u_{a0}} \end{cases}. \quad (8)$$

### D. Fixed points of the unforced system

Let us now reduce the system (6) to two algebraic equations with the excitatory rate $r_e$ and and the NMDA current in the excitatory population $I_{NMDAe}$ as the variables. In the next sections, we demonstrate that plotting the curves defined by these equations on the $(r_e\text{-}I_{NMDAe})$ phase plane provides a useful geometrical intuition about existence of solutions (which are defined by intersections of these curves), as well as about stability of these solutions.

Let us define the functions $\bar{V}_e(r_e)$ and $\bar{V}_i(r_i)$ in such way that, for an input $u_a$ ($a = e, i$), the following identity is true:

$$\bar{V}_a(r_a^{ss}(u_a)) = V_a^{ss}(u_a), \quad (9)$$

i.e., if the input $u_a$ moves the firing rate towards $r_a$, then the same input moves the membrane voltage towards $\bar{V}_a(r_a)$. From (7), it follows that:

$$\bar{V}_a(r_a) = V_{a0} + \dfrac{c_{Va}}{c_{ra}}(r_a - r_{a0}). \quad (10)$$

Next, let us find the state of the fast subsystem $(r_e, r_i, V_e, V_i)$ towards which it converges at the fast time scale for a given pair of fixed values of the slow variables $I_{NMDAe}, I_{NMDAi}$ (we denote this state as $(r_e^*, r_i^*, V_e^*, V_i^*)$). By setting the right-hand part of the first equation of (6) to zero, we get:

$$\begin{cases} r_a^*(I_{NMDAe}, I_{NMDAi}) = P_{ra} + Q_{ra}^e I_{NMDAe} + Q_{ra}^i I_{NMDAi} \\ V_a^*(I_{NMDAe}, I_{NMDAi}) = \bar{V}_a\left(r_a^*(I_{NMDAe}, I_{NMDAi})\right) \end{cases}, \qquad (11)$$

where

$$\begin{cases} Q_{re}^e = \dfrac{c_{re}(1 + c_{ri}J_{ii})}{g_{me}Q} \\ Q_{re}^i = \dfrac{-c_{re}c_{ri}J_{ei}}{g_{mi}Q} \\ Q_{ri}^e = \dfrac{c_{re}c_{ri}J_{ie}^{AMPA}}{g_{me}Q} \\ Q_{ri}^i = \dfrac{c_{ri}(1 - c_{re}J_{ee}^{AMPA})}{g_{mi}Q} \\ Q = (1 - c_{re}J_{ee}^{AMPA})(1 + c_{ri}J_{ii}) + c_{re}c_{ri}J_{ei}J_{ie}^{AMPA} \\ P_{re} = r_{e0} - Q_{re}^e I_{NMDAe,0} - Q_{re}^i I_{NMDAi,0} \\ P_{ri} = r_{i0} - Q_{ri}^e I_{NMDAe,0} - Q_{ri}^i I_{NMDAi,0} \end{cases}. \qquad (12)$$

Now we can conclude that the steady-state firing rates and NMDA-currents should satisfy:

$$\begin{cases} I_{NMDAa} = I_{NMDAa}^{ss}\left(r_e, \bar{V}_a(r_a)\right) \\ r_a = r_a^*(I_{NMDAe}, I_{NMDAi}) \end{cases}. \qquad (13)$$

Using (11), let us express $r_i$ and $I_{NMDAi}$ from (13) as functions of $r_e$ and $I_{NMDAe}$:

$$\begin{cases} \bar{r}_i(r_e, I_{NMDAe}) = \dfrac{Q_{ri}^i}{Q_{re}^i}\left[r_e - \left(P_{re} + Q_{re}^e I_{NMDAe}\right)\right] + \left(P_{ri} + Q_{ri}^e I_{NMDAe}\right) \\ \bar{I}_{NMDAi}(r_e, I_{NMDAe}) = I_{NMDAi}^{ss}\left(r_e, \bar{V}_i(\bar{r}_i(r_e, I_{NMDAe}))\right) \end{cases}, \qquad (14)$$

Finally, in order obtain self-consistent equations for $r_e$ and $I_{NMDAe}$, let us put (14) back into (13):

$$\begin{cases} r_e = P_{re} + Q_{re}^e I_{NMDAe} + Q_{re}^i \bar{I}_{NMDAi}(r_e, I_{NMDAe}) \\ I_{NMDAe} = I_{NMDAe}^{ss}\left(r_e, \bar{V}_e(r_e)\right) \end{cases}. \qquad (15)$$

As we mentioned previously, the first and the second equations of (15) define two curves on the $(r_e, I_{NMDAe})$-plane, intersections of which correspond to the fixed points of the system (6). Further in this text, we will refer to these curves as the $r_e$-curve and the $I_{NMDAe}$-curve, respectively. We should note that these curves are not nullclines, although they intersect at the fixed points of the system.

### E. Analysis of the linearized system with external periodic input: forced oscillations

Our goal is to understand under what condition external periodic forcing may change the mean activity of the neuronal populations in our model . We start by analyzing the linearized system (6) with periodic external forcing. Here we derive the amplitude and phase relations between the external periodic signal and the forced oscillations of the excitatory and inhibitory populations.

Let us apply an external periodic forcing to the system (6), with the circular frequency $\omega$ and the complex-valued amplitudes of the oscillatory inputs to the E- and I-populations equal to $h_e^A$ and $h_i^A$, respectively:

$$\begin{cases} \dfrac{dr_a}{dt} = \dfrac{1}{\tau_{ra}} \left( r_a^{ss}(u_a^{osc}(r_e, r_i, I_{NMDAa}, t)) - r_a \right) \\ \dfrac{dV_a}{dt} = \dfrac{1}{\tau_{Va}} \left( V_a^{ss}(u_a^{osc}(r_e, r_i, I_{NMDAa}, t)) - V_a \right) \\ \dfrac{dI_{NMDAa}}{dt} = \dfrac{1}{\tau_{NMDA}} \left( I_{NMDAa}^{ss}(\mathrm{Re}(r_e), \mathrm{Re}(V_a)) - I_{NMDAa} \right) \\ u_a^{osc}(r_e, r_i, I_{NMDAa}, t) = J_{ae}^{AMPA} r_e - J_{ai} r_i + I_{NMDAa}/g_{ma} + h_a + h_a^A e^{i\omega t} \end{cases} \quad (16)$$

Using the time scale separation, we can assume that: (1) the slow variables $I_{NMDAe}$ and $I_{NMDAi}$ do not get entrained by the external forcing (which is reasonable when the forcing frequency is not extremely low); and (2) for a certain combination $(I_{NMDAe}, I_{NMDAi})$, the dynamics of the fast variables $r_e, r_i, V_e, V_i$ could be represented as harmonic oscillations about the attracting fixed point of the unforced fast subsystem $(r_e^*, r_i^*, V_e^*, V_i^*)$ (see (11)). Thus, if we denote the complex-valued amplitudes of the forced firing rate oscillations as $r_e^A$, $r_i^A$, and the complex-valued amplitudes of the forced membrane voltage oscillations as $V_e^A$, $V_i^A$, then the dynamics of the fast subsystem could be expressed as follows:

$$\begin{cases} r_a(t) = r_a^*(I_{NMDAe}, I_{NMDAi}) + r_a^A e^{i\omega t} \\ V_a(t) = V_a^*(I_{NMDAe}, I_{NMDAi}) + V_a^A e^{i\omega t} \end{cases}. \quad (17)$$

The dynamics of $(r_e, r_i)$, defined by the system (16), could be expressed in the matrix form:

$$\dfrac{d}{dt}\begin{pmatrix} r_e \\ r_i \end{pmatrix} = M_r \begin{pmatrix} r_e \\ r_i \end{pmatrix} + \begin{pmatrix} \left[ c_{re}\left( I_{NMDAe}/g_{me} + h_e + h_e^A e^{i\omega t} - u_{e0} \right) + r_{e0} \right]/\tau_{re} \\ \left[ c_{ri}\left( I_{NMDAi}/g_{mi} + h_i + h_i^A e^{i\omega t} - u_{i0} \right) + r_{i0} \right]/\tau_{ri} \end{pmatrix}, \quad (18)$$

where

$$M_r = \begin{pmatrix} \dfrac{c_{re} J_{ee}^{AMPA} - 1}{\tau_{re}} & \dfrac{-c_{re} J_{ei}}{\tau_{re}} \\ \dfrac{c_{ri} J_{ie}^{AMPA}}{\tau_{ri}} & \dfrac{-(c_{ri} J_{ii} + 1)}{\tau_{ri}} \end{pmatrix}. \quad (19)$$

Now we put the expression for $r_a(t)$ from (17) into (18) and take into account the fact that $r_a = r_a^*(I_{NMDAe}, I_{NMDAi})$ is the solution of (18) when $h_e^A = h_i^A = 0$. After all cancellations, we get the following equation for the amplitudes $(r_e^A, r_i^A)$:

$$i\omega \begin{pmatrix} r_e^A \\ r_i^A \end{pmatrix} = M_r \begin{pmatrix} r_e^A \\ r_i^A \end{pmatrix} + \begin{pmatrix} h_e^A c_{re}/\tau_{re} \\ h_i^A c_{ri}/\tau_{ri} \end{pmatrix}. \quad (20)$$

Using (20), we can find the external amplitudes $(h_e^A, h_i^A)$ that would produce the forced oscillations of the firing rate with the amplitudes $(r_e^A, r_i^A)$:

$$h_a^A = \dfrac{1}{c_{ra}} r_a^A (i\omega \tau_{ra} + 1) - J_{ae}^{AMPA} r_e^A + J_{ai} r_i^A. \quad (21)$$

As the next step, we want to express the oscillatory part of the total input $u_a^{osc}$ as a function of $(r_e^A, r_i^A)$. Let us put the expression $r_a(t) = r_a^*(I_{NMDAe}, I_{NMDAi}) + r_a^A e^{i\omega t}$ from (17) and the expression for $h_a^A$ from (21) into the expression for $u_a^{osc}$ from (16):

$$u_a^{osc}\left(r_e^* + r_e^A e^{i\omega t}, r_i^* + r_i^A e^{i\omega t}, I_{NMDAa}, t\right) = u_a\left(r_e^*, r_i^*, I_{NMDAa}\right) + \frac{1}{c_{ra}} r_a^A \left(1 + i\omega \tau_{ra}\right) e^{i\omega t}, \tag{22}$$

where $u_a\left(r_e^*, r_i^*, I_{NMDAa}\right)$ is the total input received under a certain state of the slow subsystem, given that the fast subsystem converged to its steady-state (in the absence of the forcing). Here, we omitted the dependence of $r_e^*, r_i^*$ on $I_{NMDAe}, I_{NMDAi}$ for notational simplicity.

Now let us find the how the amplitude of the forced voltage and the forced firing rate are related $V_a^A$ and $r_a^A$. Let us put (22) and the expression $V_a(t) = V_a^*(I_{NMDAe}, I_{NMDAi}) + V_a^A e^{i\omega t}$ from (17) into the equation for $V_a$ from (16). Using the expression (7) for $V_a^{ss}$, and taking into account that $V_a^* = V_a^{ss}\left(u(r_e^*, r_i^*, I_{NMDAa})\right)$, we get:

$$i\omega V_a^A = \frac{1}{\tau_{Va}}\left(r_a^A \frac{c_{Va}}{c_{ra}}(1 + i\omega\tau_{ra}) - V_a^A\right). \tag{23}$$

Solving (23) with respect to $V_a^A$ yields:

$$V_a^A = r_a^A \frac{c_{Va}(1 + i\omega\tau_{ra})}{c_{ra}(1 + i\omega\tau_{Va})}. \tag{24}$$

In our analysis, we found it convenient to parametrize our system by $r_e^A, r_i^A$, instead of $h_e^A, h_i^A$. The latter pair of parameters could be derived using (21). The relation (24) will be required in the next section, where we express the forcing-induced shift of the time-averaged equilibrium.

**F. System with external periodic input: forced shift of the time-averaged equilibrium**

In this section, we analyze the influence of the forced oscillations of the fast $(r_e, r_i, V_e, V_i)$-subsystem on the dynamics of the slow $(I_{NMDAe}, I_{NMDAi})$-subsystem due to its non-linearity.

First, let us introduce short notations for the functions that govern the dynamics of the NMDA-currents (see (6) and (2)):

$$G_a\left(r_e, V_a, I_{NMDAa}\right) = \frac{1}{\tau_{NMDA}}\left(I_{NMDAa}^{ss}(r_e, V_a) - I_{NMDAa}\right). \tag{25}$$

Now let us apply the time scale separation and rewrite the equations for the NMDA dynamics without periodic forcing, substituting the fast variables $r_a, V_a$ in (25) by the values $r_a^*, V_a^*$, towards which these variables converge at the fast time scale:

$$\begin{cases} \dfrac{dI_{NMDAa}}{dt} = G_a^*(I_{NMDAe}, I_{NMDAi}) \\ G_a^*(I_{NMDAe}, I_{NMDAi}) = G_a\left(r_e^*(I_{NMDAe}, I_{NMDAi}), V_a^*(I_{NMDAe}, I_{NMDAi}), I_{NMDAa}\right) \end{cases}, \tag{26}$$

where $r_a^*, V_a^*$ are given by (11).

To account for the effect of the external forcing, we write the equations for the slow dynamics with $r_a, V_a$ not equal to $r_a^*, V_a^*$ (as in (26)), but oscillating about $r_a^*, V_a^*$ (see (17):

$$\begin{cases} \dfrac{dI_{NMDAa}}{dt} = G_a^{osc}(I_{NMDAe}, I_{NMDAi}, t) \\ G_a^{osc}(I_{NMDAe}, I_{NMDAi}, t) = G_a\left(r_e^* + |r_e^A|\cos(\omega t + \psi_{ra}), V_a^* + |V_a^A|\cos(\omega t + \psi_{Va}), I_{NMDAa}\right) \end{cases}, \tag{27}$$

where $\psi_{ra} = \arg r_a^A$, $\psi_{Va} = \arg V_a^A$.

In order to estimate the effect of the forced oscillations on the slow dynamics, we apply the averaging method (Strogatz, 2015). We introduce the new slow variables $\tilde{I}_{NMDAe}$ and $\tilde{I}_{NMDAi}$ that describe the values of $I_{NMDAe}$ and $I_{NMDAi}$, respectively, averaged over the forcing period $T = 2\pi/\omega$. From the time scale separation, it follows that $T$ is much smaller than the characteristic time of the slow dynamics. Consequently, the dynamics of the time-averaged variables $\tilde{I}_{NMDAe}, \tilde{I}_{NMDAi}$ could be expressed as follows:

$$\frac{d\tilde{I}_{NMDAa}}{d\tilde{t}} = \frac{1}{T}\int_{\tilde{t}}^{\tilde{t}+T} G_a^{osc}(I_{NMDAe}, I_{NMDAi}, t)dt. \qquad (28)$$

Let us write an explicit expression for $G_a^{osc}$, using (25) and (27):

$$G_a^{osc}(I_{NMDAe}, I_{NMDAi}, t) = $$
$$= \frac{1}{\tau_{NMDA}}\left(J_{ae}^{NMDA} g_{NMDA}\left(V_a^* + |V_a^A|\cos(\omega t + \psi_{Va})\right)\left(r_e^* + |r_e^A|\cos(\omega t + \psi_{ra})\right) - I_{NMDAa}\right). \qquad (29)$$

In order to analytically estimate the integral in (28), let us expand $g_{NMDA}$ about $V_a^*$:

$$\begin{cases} g_{NMDA}(V_a) = \sum_{k=0}^{\infty} \frac{1}{k!} g_{ak}\left(V_a - V_a^*\right)^k \\ g_{ak} = \left.\frac{dg_{NMDA}(V_a)}{dV_a}\right|_{V_a^*} \end{cases}. \qquad (30)$$

From the expansion (30), we get:

$$g_{NMDA}\left(V_a^* + |V_a^A|\cos(\omega t + \psi_{Va})\right)\left(r_e^* + |r_e^A|\cos(\omega t + \psi_{ra})\right) = $$
$$= r_e^* g_{NMDA}(V_a^*) + |r_e^A| g_{NMDA}(V_a^*)\cos(\omega t + \psi_{ra}) + r_e^* |V_a^A| g'_{NMDA}(V_a^*)\cos(\omega t + \psi_{Va}) +$$
$$+ |r_e^A||V_a^A| g'_{NMDA}(V_a^*)\cos(\omega t + \psi_{Va})\cos(\omega t + \psi_{ra}) + \qquad (31)$$
$$+ \frac{1}{2} r_e^* |V_a^A|^2 g''_{NMDA}(V_a^*)\cos^2(\omega t + \psi_{Va}) + o\left(|r_e^A|^2\right)$$

where we assume that $|r_e^A|, |r_i^A|, |V_e^A|, |V_i^A|$ have the same order of magnitude.

Now we put (31) into (29), omitting the higher-order terms, and calculate the integral in (28). As we integrate over the period of oscillations, all time-dependent cosine terms cancel out, and we get an autonomous equation:

$$\begin{cases} \dfrac{d\tilde{I}_{NMDAa}}{d\tilde{t}} = \tilde{G}_a\left(\tilde{I}_{NMDAe}, \tilde{I}_{NMDAi}\right) \\ \tilde{G}_a\left(\tilde{I}_{NMDAe}, \tilde{I}_{NMDAi}\right) = G_a^*\left(\tilde{I}_{NMDAe}, \tilde{I}_{NMDAi}\right) + \dfrac{D_a\left(\tilde{r}_e, \tilde{V}_a\right)}{\tau_{NMDA}} \\ D_a\left(\tilde{r}_e, \tilde{V}_a\right) = \dfrac{1}{4} J_{ae}^{NMDA}\left[2|r_e^A||V_a^A| g'_{NMDA}\left(\tilde{V}_a\right)\cos\varphi_a + \tilde{r}_e |V_a^A|^2 g''_{NMDA}\left(\tilde{V}_a\right)\right] \end{cases}, \qquad (32)$$

where $\varphi_a = \psi_{Va} - \psi_{re}$ is the phase lag between the voltage and the firing rate oscillations, and $\tilde{r}_a, \tilde{V}_a$ are newly introduced variables such that:

$$\begin{cases} \tilde{r}_a \equiv r_a^*\left(\tilde{I}_{NMDAe}, \tilde{I}_{NMDAi}\right) \\ \tilde{V}_a \equiv V_a^*\left(\tilde{I}_{NMDAe}, \tilde{I}_{NMDAi}\right) \end{cases}. \qquad (33)$$

We can consider $\left(\tilde{r}_e, \tilde{r}_i, \tilde{V}_e, \tilde{V}_i\right)$ as a slowly moving "center", around which the fast subsystem oscillates under the external forcing. From now on, we will refer to the $(\tilde{I}_{NMDAe}, \tilde{I}_{NMDAi})$-system described by (32), together with the variables $\tilde{r}_e, \tilde{r}_i, \tilde{V}_e, \tilde{V}_i$ that functionally depend on $\tilde{I}_{NMDAe}, \tilde{I}_{NMDAi}$, as the time-averaged forced system.

The first term in the square brackets in (32) (containing $|r_e^A||V_a^A|$) reflects the fact that NMDA-current depends both on the presynaptic firing rate and the postsynaptic voltage, while the second term (containing $|V_a^A|^2$) reflects the fact that NMDA-current depends non-linearly on the postsynaptic voltage. Our numerical results (see the following sections) show that, for the selected parameters, the first term is much larger than the second one. Thus, the forcing-induced shift of the time-averaged equilibrium is mainly related to the joint effect of the presynaptic firing rate oscillations and the postsynaptic voltage oscillations, occurring with a small phase lag between them.

We parametrize our system by $r_e^A$ and $r_i^A$, so we want to express $D_e, D_i$ in the terms of $r_e^A, r_i^A$. In order to do this, we derive from (24) the amplitudes $|V_e^A|, |V_i^A|$:

$$|V_a^A| = |r_a^A| \frac{c_{Va}(1+\omega^2 \tau_{ra}^2)}{c_{ra}(1+\omega^2 \tau_{Va}^2)}, \tag{34}$$

and the cosines of the phase lags:

$$\begin{cases} \cos\varphi_e = \cos\arg \dfrac{V_e^A}{r_e^A} = \dfrac{1+\omega^2 \tau_{re}\tau_{Ve}}{\sqrt{(1+\omega^2 \tau_{Ve}^2)(1+\omega^2 \tau_{re}^2)}} \\ \cos\varphi_i = \cos\arg \dfrac{V_i^A}{r_e^A} = \dfrac{(1+\omega^2 \tau_{ri}\tau_{Vi})\cos\gamma_r - \omega(\tau_{ri}-\tau_{Vi})\sin\gamma_r}{\sqrt{(1+\omega^2 \tau_{Vi}^2)(1+\omega^2 \tau_{ri}^2)}} \end{cases}, \tag{35}$$

where $\gamma_r$ is the angle between $r_e^A$ and $r_i^A$.

Taking into account (25) and (26), the steady-states of (32) should satisfy:

$$\begin{cases} \tilde{I}_{NMDAa} = I_{NMDAa}^{ss}(\tilde{r}_e, \bar{V}_a(\tilde{r}_a)) + D_a(\tilde{r}_e, \bar{V}_a(\tilde{r}_a)) \\ \tilde{r}_a = r_a^*(\tilde{I}_{NMDAe}, \tilde{I}_{NMDAi}) \end{cases}, \tag{36}$$

which is the analog of (13) in the case of the time-averaged forced system. Now let us exclude $\tilde{r}_i$ and $\tilde{I}_{NMDAi}$ from (36), similarly to (14), and get the self-consistent expression for the steady-state values of $\tilde{r}_e$ and $\tilde{I}_{NMDAe}$, which is analogous to (15):

$$\begin{cases} \tilde{r}_e = P_{re} + Q_{re}^e \tilde{I}_{NMDAe} + Q_{re}^i \bar{I}_{NMDAi}(\tilde{r}_e, \tilde{I}_{NMDAe}) + Q_{re}^i D_i(\tilde{r}_e, \bar{V}_i(\bar{r}_i(\tilde{r}_e, \tilde{I}_{NMDAe}))) \\ \tilde{I}_{NMDAe} = I_{NMDAe}^{ss}(\tilde{r}_e, \bar{V}_e(\tilde{r}_e)) + D_e(\tilde{r}_e, \bar{V}_e(\tilde{r}_e)) \end{cases}, \tag{37}$$

where $\bar{r}_i$ and $\bar{I}_{NMDAi}$ are the same function as it was introduced in (14).

Similar to the unforced case, the steady-states of the time-averaged forced system could be found as intersections of the $\tilde{r}_e$-curve and the $\tilde{I}_{NMDAe}$-curve on the $(\tilde{r}_e, \tilde{I}_{NMDAe})$-plane, which are defined by the first and the second equations of (37), respectively. Both $g'_{NMDA}$ and $g''_{NMDA}$ are positive in the physiological range of voltages. Also, in the results section we demonstrate that, for the selected parameters, $\cos\varphi_e, \cos\varphi_i$ are also positive. Consequently, both $D_e$ and $D_i$ are positive. Also, for $J_{ee}^{AMPA}$ being small enough, $Q_{re}^i < 0$. Thus, from (37) and (15), it follows that the $\tilde{I}_{NMDAe}$-curve is shifted upwards relatively to the $I_{NMDAe}$-curve, and the $\tilde{r}_e$-curve is shifted to the left relatively to the $r_e$-curve. In the following sections, we will demonstrate that in case of $J_{ie}^{AMPA} = 0$, such character of the $\tilde{I}_{NMDAe}$- and the $\tilde{r}_e$-curve shifts implies that the external forcing always increases the mean firing rates.

Further in this paper, we denote the fixed point of the time-averaged forced system that corresponds to the fixed point $(r_{e0}, r_{i0}, V_{e0}, V_{i0}, I_{NMDAe0}, I_{NMDAi0})$ of the initial system as $(\tilde{r}_{e0}, \tilde{r}_{i0}, \tilde{V}_{e0}, \tilde{V}_{i0}, \tilde{I}_{NMDAe0}, \tilde{I}_{NMDAi0})$.

**G. Stability analysis of the unforced and the periodically forced systems**

Let us now investigate the stability conditions for the systems with and without the periodic forcing. We want to identify conditions under which the periodic forcing changes the time-averaged activity without destabilizing it.

For the unforced system, we can apply time scale separation and analyze stability of the fast and the slow subsystems independently. The fast $(r_e, r_i, V_e, V_i)$ subsystem is stable when all the eigenvalues of the matrix $M_r$ (given by (19)) have negative real parts. Stability of the slow $(I_{NMDAe}, I_{NMDAi})$ subsystem of the unforced system could be determined by analyzing (26). In the case of $J_{ie}^{NMDA} = 0$, the slow subsystem is stable, if:

$$\left. \frac{dG_e^*(I_{NMDAe})}{dI_{NMDAe}} \right|_{I_{NMDAe0}} < 0. \tag{38}$$

In the case of $J_{ie}^{NMDA} \neq 0$, one should consider the matrix:

$$M_{NMDA} = \left. \begin{pmatrix} \dfrac{dG_e^*}{dI_{NMDAe}} & \dfrac{dG_e^*}{dI_{NMDAi}} \\ \dfrac{dG_i^*}{dI_{NMDAe}} & \dfrac{dG_i^*}{dI_{NMDAi}} \end{pmatrix} \right|_{I_{NMDAe0}, I_{NMDAi0}}. \tag{39}$$

The slow subsystem is stable if all the eigenvalues of $M_{NMDA}$ have negative real parts.

Applying the time scale separation to the forced system, its asymptotic dynamics could be considered as forced oscillations of the fast $(r_e, r_i, V_e, V_i)$ subsystem around the point $(\tilde{r}_{e0}, \tilde{r}_{i0}, \tilde{V}_{e0}, \tilde{V}_{i0})$ that functionally depends on the equilibrium $(\tilde{I}_{NMDAe0}, \tilde{I}_{NMDAi0})$ of the time-averaged slow subsystem $(\tilde{I}_{NMDAe}, \tilde{I}_{NMDAi})$. As the matrix $M_r$ is constant, stability of the fast subsystem does not depend on the state of the slow subsystem. Consequently, if the fast subsystem of the unforced system is stable, it is guaranteed that the forced oscillations will have a finite amplitude (at least in the case when this amplitude is small enough, so we do not need to take fixed points other than $(\tilde{I}_{NMDAe0}, \tilde{I}_{NMDAi0})$ into consideration).

The stability conditions for the time-averaged forced slow subsystem are similar to (38) and (39). For $J_{ie}^{NMDA} = 0$ we get:

$$\left. \frac{d\tilde{G}_e(\tilde{I}_{NMDAe})}{d\tilde{I}_{NMDAe}} \right|_{\tilde{I}_{NMDAe0}} < 0, \tag{40}$$

and for $J_{ie}^{NMDA} \neq 0$ we require the real parts of the eigenvalues of the following matrix to be negative:

$$\tilde{M}_{NMDA} = \left. \begin{pmatrix} \dfrac{d\tilde{G}_e}{d\tilde{I}_{NMDAe}} & \dfrac{d\tilde{G}_e}{d\tilde{I}_{NMDAi}} \\ \dfrac{d\tilde{G}_i}{d\tilde{I}_{NMDAe}} & \dfrac{d\tilde{G}_i}{d\tilde{I}_{NMDAi}} \end{pmatrix} \right|_{\tilde{I}_{NMDAe0}, \tilde{I}_{NMDAi0}}. \tag{41}$$

We should note that the time-averaged forced system may have no equilibria (i.e. the $\tilde{r}_e$-curve and the $\tilde{I}_{NMDAe}$-curve do not intersect). In this case, the conditions (40) and (41) are, obviously, not applicable, and the system diverges.

Let us transform the stability conditions (38) and (40) to a form that is suitable for further geometrical analysis. We can use the expressions (26) and (25) for $G_e^*$ and $G_e$ to explicitly write down the derivative $dG_e^*/dI_{NMDAe}$, then apply the chain rule: $d/dI_{NMDAe} = (d/dr_e) \cdot (dr_e^*/dI_{NMDAe})$, and take into account that, according to (11), $dr_e^*/dI_{NMDAe} = Q_{re}^e$. The same procedure could be applied to $d\tilde{G}_e/d\tilde{I}_{NMDAe}$

, with the exception that $\tilde{G}_e$ differs from $G_e^*$ by the function $D_e$. As the result, the derivatives in (38) and (40) could be expressed as:

$$\begin{cases} \dfrac{dG_e^*(I_{NMDAe})}{dI_{NMDAe}}\bigg|_{I_{NMDAe0}} = \dfrac{1}{\tau_{NMDA}}\left[-1 + \dfrac{dI_{NMDAe}^{ss}(r_e, \bar{V}_e(r_e))}{dr_e}\bigg|_{r_{e0}} \cdot Q_{re}^e \right] \\ \dfrac{d\tilde{G}_e(\tilde{I}_{NMDAe})}{d\tilde{I}_{NMDAe}}\bigg|_{\tilde{I}_{NMDAe0}} = \dfrac{1}{\tau_{NMDA}}\left[-1 + \dfrac{d}{d\tilde{r}_e}\left(I_{NMDAe}^{ss}(\tilde{r}_e, \bar{V}_e(\tilde{r}_e)) + D_e(\tilde{r}_e, \bar{V}_e(\tilde{r}_e))\right)\bigg|_{\tilde{r}_{e0}} \cdot Q_{re}^e \right] \end{cases}. \quad (42)$$

For $J_{ee}^{AMPA}$ being small enough, $Q_{re}^e > 0$. Consequently, taking into account (42), the stability conditions (38) and (40) (that correspond to the case $\tilde{J}_{ie}^{NMDA} = 0$) could be rewritten as follows:

$$\begin{cases} \dfrac{dI_{NMDAe}^{ss}(r_e, \bar{V}_e(r_e))}{dr_e}\bigg|_{r_{e0}} < \dfrac{1}{Q_{re}^e} \\ \dfrac{d}{d\tilde{r}_e}\left(I_{NMDAe}^{ss}(\tilde{r}_e, \bar{V}_e(\tilde{r}_e)) + D_e(\tilde{r}_e, \bar{V}_e(\tilde{r}_e))\right)\bigg|_{\tilde{r}_{e0}} < \dfrac{1}{Q_{re}^e} \end{cases}. \quad (43)$$

We further discuss (43) in the results section, where we consider the system with $\tilde{J}_{ie}^{NMDA} = 0$; we also provide geometrical interpretation for these conditions in terms of positions of the $r_e, \tilde{r}_e, I_{NMDAe}, \tilde{I}_{NMDAe}$-curves on the phase plane. In the subsequent results sections, where we consider the case of $\tilde{J}_{ie}^{NMDA} \neq 0$, we check the stability conditions (39) and (41) numerically.

## III. PARAMETER SELECTION AND VALIDATION

### A. Parameter selection

In this section, we briefly describe the selection of the primary parameters of our model, as well as calculation of the parameters derived from the primary ones.

In this work, we are mainly interested in the effect of the recurrent synaptic weights on the system's ability to change its time-averaged steady state under external forcing. To separate this effect from possible influence of the unforced steady state change caused by changes in the synaptic weights, we a priori select the steady state of the unforced system and then vary the synaptic weights, automatically re-tuning the external input strengths for each combination of the weights in such way that the pre-selected steady-state is kept constant. More precisely, we fixed (i.e. set a priori) the total mean inputs $u_{e0}, u_{i0}$ and the total input variances $\sigma_{e0}^2, \sigma_{i0}^2$, and calculated the external mean inputs $h_e, h_i$ and the external input variances $\sigma_{xe}^2, \sigma_{xi}^2$ as functions of the total input parameters and of the synaptic weights.

The same was done for the oscillatory part of the external input: we fixed the amplitudes $|r_e^A|, |r_i^A|$ of the forced oscillations of the excitatory and inhibitory populations, as well as the phase lag $\gamma_r$ between them, and then we derived the amplitudes $|h_e^A|, |h_i^A|$ of external oscillations delivered to the populations (and the corresponding phases) as functions of $|r_e^A|, |r_i^A|, \gamma_r$ and of the synaptic weights.

The linearization coefficients $c_{ra}, c_{Va}$ that define the slopes of $r_a^{ss,0}, V_a^{ss,0}$ at $u_{a0}$ given $\sigma_{a0}$ (see (8)) were derived from simulations of individual leaky integrate-and-fire (LIF) neurons (one excitatory and one inhibitory). These LIF neurons were described by the following parameters: the membrane time constant $\tau_{ma}$, the resting potential $E_{La}$, the spiking threshold $V_a^{th}$, the reset voltage $V_a^r$. We performed simulations for values of $u_a$ close to $u_{a0}$, obtained numerical estimations of $r_a^{ss,0}(u_a), V_a^{ss,0}(u_a)$ for each of them, and then calculated $c_{ra}, c_{Va}$ using these estimations. Also, we performed similar simulations for $u_a = u_{a0}$ and calculated CV values $CV_{a0}$ from the obtained spike trains.

In order to find population time constants $\tau_{ra}$ and $\tau_{Va}$, we considered a population of uncoupled LIF neurons with the corresponding parameters. We delivered a small input step that shifted the total input from $u_{a0}$, and measured the dynamics $r_a(t)$ and $V_a(t)$ that the network demonstrated in response to this input step. Then we fitted $r_a(t)$ and $V_a(t)$ with exponential functions, from which we obtained the constants $\tau_{ra}$ and $\tau_{Va}$.

The aforementioned procedures of parameter derivation are described in more details in the Appendix.

To pursue our analysis and make clear the role of the non-linear NMDA synaptic currents, we considered two models. The first model (Model 1) contained NMDA currents on the excitatory neurons only ($j_{ie}^{NMDA}=0$), and the second model (Model 2) contained NMDA currents both on the excitatory and the inhibitory neurons ($j_{ie}^{NMDA} \neq 0$). The parameters of both models were selected in such way that provided realistic mode of operation, as well as existence and stability of the time-averaged forced equilibrium. The synaptic weight $\tilde{J}_{ie}^{AMPA}$ was decreased in the Model 2 (compared to the Model 1) in order to demonstrate that the fast subsystem could be made more excitable in the presence of the slow NMDA inhibition without loss of stability (which leads to a more pronounced forcing-induced shift of the mean excitatory firing rate). All the primary parameters of the Models 1 and 2 models, except of $j_{ie}^{AMPA}$ and $j_{ie}^{NMDA}$, are the same.

During the selection of parameters, we used the following convenient property of our system: one can note that the $\tilde{I}_{NMDAe}$-curve does not depend on the fast synaptic weights $J_{ee}^{AMPA}$, $J_{ie}^{AMPA}$, $J_{ei}$, $J_{ii}$. Thus, we were able to control the fixed point of the time-averaged forced system by changing the fast synaptic weights (at the same time automatically adjusting the external inputs, as described above). Change of the fast weights caused "rotation" of the $\tilde{r}_e$-curve around the unforced equilibrium $(r_{e0}, I_{NMDAe0})$, not affecting the $\tilde{I}_{NMDAe}$-curve, so the position of the time-averaged forced equilibrium $(\tilde{r}_{e0}, \tilde{I}_{NMDAe0})$ was easy to predict.

The primary parameters are listed in the Table 1; the derived parameters, as well as characteristics of equilibria are listed in the Table 2.

Table 1. Primary parameters

| | | | | | |
|---|---|---|---|---|---|
| $K_e$ | 800 | $\tau_{me}$ | 20 ms | $u_{e0}$ | -67.25 mV |
| $K_i$ | 200 | $\tau_{mi}$ | 10 ms | $u_{i0}$ | -65.25 mV |
| $\tau_{NMDA}$ | 200 ms | $g_{me}=g_{mi}$ | 100 μS/cm² | $\sigma_{e0}$ | 14.2 mV |
| $j_{ee}^{AMPA}$ | 0 mV | $V_e^r=V_i^r$ | -70 mV | $\sigma_{i0}$ | 13 mV |
| $j_{ee}^{NMDA}$ | 0.15 μA/cm² | $V_e^{th}=V_i^{th}$ | -55 mV | $\omega/2\pi$ | 20 Hz |
| $\tilde{J}_{ei}$ | -1 mV | $E_{Le}=E_{Li}$ | -70 mV | $|r_e^A|$ | 10 Hz |
| $\tilde{J}_{ii}$ | 0 mV | | | $|r_i^A|$ | 5 Hz |
| | | | | $\gamma_r$ | 0 |
| **Model 1** | $j_{ie}^{AMPA}$ | 0.117 mV | **Model2** | $j_{ie}^{AMPA}$ | 0.06 mV |
| | $j_{ie}^{NMDA}$ | 0 μA/cm² | | $j_{ie}^{NMDA}$ | 0.003 μA/cm² |

Table 2. Derived parameters and characteristics of equilibria

| | | | | | |
|---|---|---|---|---|---|
| $r_{e0}$ | 24.2 Hz | $V_{e0}$ | -75.06 mV | $CV_{e0}$ | 1.1 |
| $r_{i0}$ | 29.5 Hz | $V_{i0}$ | -69.93 mV | $CV_{i0}$ | 0.9 |
| $c_{re}$ | 1921.2 | $\tau_{re}$ | 6.2 ms | $|V_e^A|$ | 2.37 mV |
| $c_{ri}$ | 3528.9 | $\tau_{ri}$ | 2.8 ms | $|V_i^A|$ | 0.60 mV |
| $c_{Ve}$ | 0.506 | $\tau_{Ve}$ | 7.9 ms | $\cos\varphi_e$ | 0.993 |
| $c_{Vi}$ | 0.454 | $\tau_{Vi}$ | 4.4 ms | $\cos\varphi_i$ | 0.986 |
| **Model 1** | | | | | |
| $h_{xe}$ | -69.79 mV | $\sigma_{xe}$ | 11.94 mV | $I_{NMDAe0}$ | 19.07 μA/cm² |
| $h_{xi}$ | -17.86 mV | $\sigma_{xi}$ | 9.12 mV | $I_{NMDAi0}$ | 0 μA/cm² |
| $|h_e^A|$ | 25.53 mV | $\arg h_e^A$ | 0.16 | | |
| $|h_i^A|$ | 7.96 mV | $\arg h_i^A$ | 3.08 | | |
| $\tilde{r}_{e0}$ | 27 Hz | $\tilde{V}_{e0}$ | -74.31 mV | $\tilde{I}_{NMDAe0}$ | 23 μA/cm² |
| $\tilde{r}_{i0}$ | 39 Hz | $\tilde{V}_{i0}$ | -68.71 mV | $\tilde{I}_{NMDAi0}$ | 0 μA/cm² |
| **Model 2** | | | | | |
| $h_{xe}$ | -69.79 mV | $\sigma_{xe}$ | 11.94 mV | $I_{NMDAe0}$ | 19.07 μA/cm² |
| $h_{xi}$ | -12.03 mV | $\sigma_{xi}$ | 9.17 mV | $I_{NMDAi0}$ | 0.52 μA/cm² |
| $|h_e^A|$ | 25.53 mV | $\arg h_e^A$ | 0.16 | | |
| $|h_i^A|$ | 3.42 mV | $\arg h_i^A$ | 3.00 | | |
| $\tilde{r}_{e0}$ | 36.8 Hz | $\tilde{V}_{e0}$ | -71.73 mV | $\tilde{I}_{NMDAe0}$ | 36.27 μA/cm² |
| $\tilde{r}_{i0}$ | 70.9 Hz | $\tilde{V}_{i0}$ | -64.61 mV | $\tilde{I}_{NMDAi0}$ | 1.08 μA/cm² |

### B. Parameter validation

In this section, we demonstrate that the parameters we selected are physiologically plausible and lead to realistic behavior of the system.

Post-synaptic potentials produced by activation of the fast receptors are given by the synaptic weights $\tilde{J}_{ee}^{AMPA}, \tilde{J}_{ie}^{AMPA}, \tilde{J}_{ei}, \tilde{J}_{ii}$. Post-synaptic potentials produced by the NMDA receptors activation are equal to 1.16 mV for the E-E connections and to 0.026 mV for the E-I connection. All these value lie in the physiological range roughly below 1.5 mV.

In our model, we define the external inputs by their mean values $h_{xa}$ and standard deviations $\sigma_{xa}$. However, the inputs in real networks are represented not by continuous signal but by spike trains. Let us assume that neurons from a population $a$ receive $K_{ae}^x$ excitatory and $K_{ai}^x$ inhibitory external inputs which have the synaptic weights equal to $J_{ae}^x, J_{ai}^x$ and the presynaptic firing rates equal to $r_{ae}^x, r_{ai}^x$, respectively. We want to check whether it is possible to select reasonable values for the aforementioned parameters of the external inputs that provide the values of $h_{xa}$ and $\sigma_{xa}$ that we used in the model. One of the appropriate combinations of these parameters is the following: $K_{ee}^x = K_{ie}^x = 4000$, $K_{ei}^x = K_{ii}^x = 1000$, $r_{ee}^x = r_{ie}^x = 2.5\,\text{Hz}$, $r_{ei}^x = r_{ii}^x = 5\,\text{Hz}$, $J_{ee}^x = 0.44\,\text{mV}$, $J_{ei}^x = 1.57\,\text{mV}$, $J_{ie}^x = 0.62\,\text{mV}$, $J_{ii}^x = 1.6\,\text{mV}$. We suggest that these parameters fall into physiologically reasonable range of values.

As we use linearized versions of the gain functions in our model, the firing rates could, in principle, become negative, which is non-sense. However, we fix the steady-state and the amplitude of the induced oscillations in such way that it does not happen. Specifically, the minimal firing rate for the excitatory / inhibitory population during the oscillations ($\tilde{r}_{a0} - |r_a^A|$) equals to 26.8 Hz / 65.9 Hz in the Model A, and 17 Hz / 34 Hz in the Model B.

The values of CV (numerically obtained from the simulations of individual LIF neurons) at the unforced steady-state ($r_{e0}, r_{i0}, V_{e0}, V_{i0}, I_{NMDAe0}, I_{NMDAi0}$) are equal to 1.1 for the excitatory neurons and 0.9 for the inhibitory neurons, which corresponds to Poisson-like spiking that is experimentally observed for most of the cortical neurons. Values of CV under external forcing are not fully tractable, as the spike trains are partially modulated by oscillations. However, we can consider a system that has the equilibrium state ($\tilde{r}_{e0}, \tilde{r}_{i0}, \tilde{V}_{e0}, \tilde{V}_{i0}, \tilde{I}_{NMDAe0}, \tilde{I}_{NMDAi0}$) in the presence of purely asynchronous input with appropriately increased mean. For the parameters of the Model A, the CV in such system equals to 1.1 for the excitatory neurons and 0.9 for the inhibitory neurons; for the parameters of the Model B, the CV values of the excitatory and inhibitory neurons are equal to 1.0 and 0.8, respectively. From this, we can conclude that the equilibrium shift produced by the external forcing does not move the system out of the physiologically plausible region of asynchronous firing.

The difference between the spiking threshold and the time-averaged forced equilibrium voltage for the excitatory / inhibitory neurons ($V_a^{th} - \tilde{V}_{a0}$) equals to 19.3 mV / 13.7 mV in the Model A and to 16.7 mV / 9.6 mV in the Model B. These values are large compared to the amplitudes $|V_a^A|$ of the forced voltage oscillations (2.37 mV for the excitatory neurons and 0.6 mV for the inhibitory neurons). Consequently, even during peaks of oscillations, the neurons are well below the spiking threshold, so spiking is driven by random fluctuations of the input and not by oscillations themselves. Furthermore, the ratio between the firing rate amplitude and its time-averaged equilibrium for the excitatory / inhibitory neurons ($|r_a^A|/\tilde{r}_{a0}$) equals to 0.37 / 0.13 in the Model A and 0.27 / 0.07 in the Model B. The values for the excitatory population (0.37 and 0.27) are larger than usually observed experimentally, however, they still suggest that the system is far from unrealistic oversynchronization and all-or-none spiking pattern driven by oscillations. In summary, the system demonstrates hallmarks of the sparsely synchronous regime characterized by moderate periodic firing rate modulation and irregular fluctuation-driven spiking of individual neurons which is believed to be typical for oscillatory cortical networks.

## IV. ANALYSIS OF SYSTEM WITH NMDA SYNAPSES ON THE EXCITATORY POPULATION ONLY (MODEL 1)

In this chapter, we analyze the system with NMDA-receptors located on the excitatory neurons only ($\tilde{J}_{NNMDAi} = 0$). First, we plot the $r_e$ - and $I_{NMDAe}$ -curves and discuss their relation to existence and stability of the equilibrium in the time-averaged forced system. Next, we confirm the predictions of the phase-plane analysis by a direct simulation of the system. Finally, we discuss relations between the forcing-induced shift of the equilibrium and the parameters of the model.

### A. Phase-plane analysis

Given the condition $\tilde{J}_{NMDAi} = 0$, the equations (15) for the $r_e$ - and $I_{NMDAe}$ -curves of the unforced system are expressed as follows:

$$\begin{cases} r_e = P_{re} + Q_{re}^e I_{NMDAe} \\ I_{NMDAe} = I_{NMDAe}^{ss}\left(r_e, \overline{V}_e(r_e)\right) \end{cases}, \quad (44)$$

and the similar equations (37) for the $\tilde{r}_e$ - and $\tilde{I}_{NMDAe}$ -curves of the time-averaged forced system could be expressed as follows:

$$\begin{cases} \tilde{r}_e = P_{re} + Q_{re}^e \tilde{I}_{NMDAe} \\ \tilde{I}_{NMDAe} = I_{NMDAe}^{ss}\left(\tilde{r}_e, \overline{V}_e(\tilde{r}_e)\right) + D_e\left(\tilde{r}_e, \overline{V}_e(\tilde{r}_e)\right) \end{cases}. \quad (45)$$

From (44) and (45), it is seen that the $r_e$ - and $\tilde{r}_e$ -curves are identical and both represented by a straight line.

In the Fig. 1(a), the phase plane is shown: the solid black line represents the $r_e/\tilde{r}_e$ -curve, the solid blue line represents the $I_{NMDAe}$ -curve, and the solid red line represents the $\tilde{I}_{NMDAe}$ -curve. The point $S_0$ is the equilibrium ($r_{e0}, I_{NMDAe0}$) of the unforced system, and $\tilde{S}_0$ is the equilibrium ($\tilde{r}_{e0}, \tilde{I}_{NMDAe0}$) of the time-averaged forced system. The unstable fixed point of the time-averaged forced system is denoted by $\tilde{S}_1$. A zoomed in region of the phase plane containing the equilibrium points is presented in the Figure 1B (this region is denoted by the grey rectangle in the Figure 1A). The dashed red line is the tangent line to the $\tilde{I}_{NMDAe}$ -curve that passes through $S_0$ (the tangency point at the $\tilde{I}_{NMDAe}$ -curve is denoted by $A$). The importance of this line is discussed below.

Now we provide a simple geometrical interpretation for stability of the slow subsystem of the unforced system. Taking a look at the stability condition given by the first inequality in (43), one can see that its right-hand part defines the slope of the $r_e/\tilde{r}_e$-curve, while the left-hand part defines the slope of the line tangent to the $I_{NMDAe}$-curve at $S_0$ (see the equations (44) for the $r_e/\tilde{r}_e$- and the $I_{NMDAe}$-curves). Consequently, the slow unforced subsystem is stable at $S_0$, if the $r_e/\tilde{r}_e$-curve (solid black line in the Figure 1B) goes steeper than the $I_{NMDAe}$-curve (solid blue line in the Figure 1B) at $S_0$. Also, from this condition it follows that, if $S_0$ is stable, then $\tilde{r}_{e0} > r_{e0}$ (as the $\tilde{I}_{NMDAe}$-curve goes above the $I_{NMDAe}$-curve), i.e. the forcing has an excitatory effect on the system.

Similarly, from the second inequality in (43) (which provides the stability condition for the time-averaged forced slow subsystem), one can see that its right-hand part, again, defines the slope of the $r_e/\tilde{r}_e$-curve, and the left-hand part defines the slope of the line tangent to the $\tilde{I}_{NMDAe}$-curve at $\tilde{S}_0$ (see the equations (45) for the $r_e/\tilde{r}_e$- and the $\tilde{I}_{NMDAe}$-curves). Thus, the time-averaged forced slow subsystem is stable at $\tilde{S}_0$, if the $r_e/\tilde{r}_e$-curve (solid black line in the Figure 1B) goes steeper than the $\tilde{I}_{NMDAe}$-curve (solid red line in the Figure 1B) at $\tilde{S}_0$.

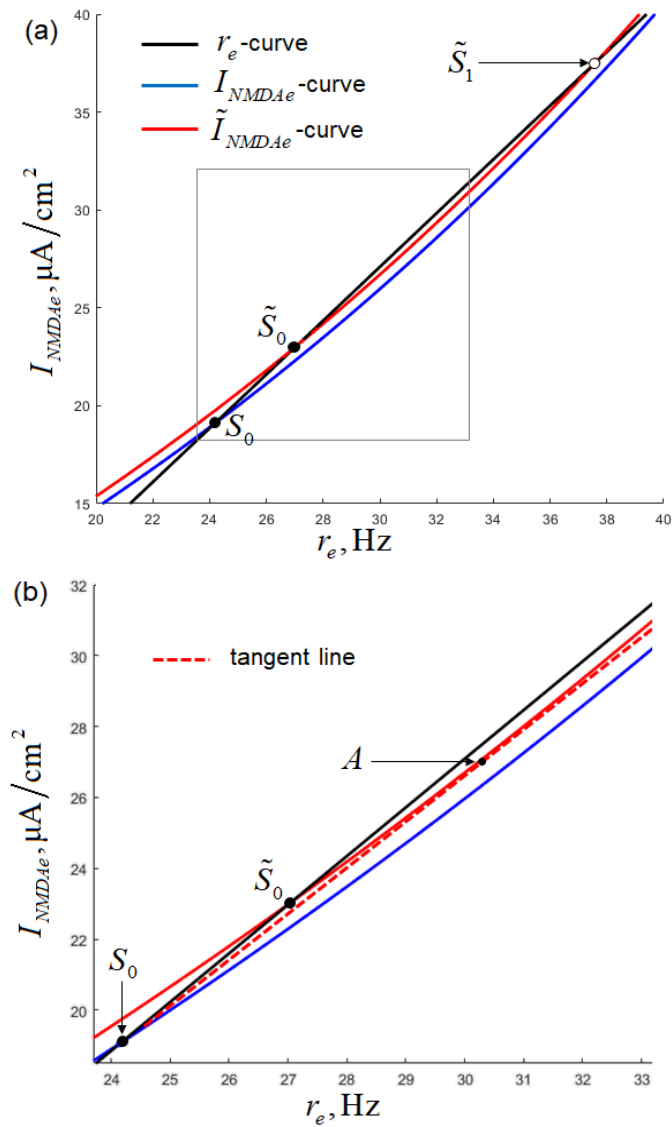

FIG 1. (a) Part of the phase plane with $r_e, \tilde{r}_e, I_{NMDAe}, \tilde{I}_{NMDAe}$-curves for Model 1. (b) Zoomed in region of interest (denoted in the panel A by the grey rectangle). Solid black line - $r_e/\tilde{r}_e$-curve; solid blue lines -

$I_{NMDAe}$-curve; solid red lines - $\tilde{I}_{NMDAe}$-curve; $S_0, \tilde{S}_0$ - equilibria of the unforced and the time-averaged forced systems, respectively; $\tilde{S}_1$ - unstable fixed point of the time-averaged forced system. Red dashed line - tangent line to the $\tilde{I}_{NMDAe}$-curve containing $S_0$; $A$ - the corresponding tangency point.

The condition for existence of the time-averaged forced equilibrium $\tilde{S}_0$ could be geometrically formulated as follows. The dynamics of the forced system is bounded only if the $r_e/\tilde{r}_e$-curve intersects with the $\tilde{I}_{NMDAe}$-curve, i.e. the $r_e/\tilde{r}_e$-curve should go steeper than the line passing through $S_0$ and tangent to the $\tilde{I}_{NMDAe}$-curve (the latter is represented by the dashed red line $(S_0 A)$ in the Figure 1B). As we make the fast subsystem more excitable (by increasing $J_{ee}^{AMPA}, J_{ii}$ or decreasing $J_{ie}^{AMPA}, J_{ei}$), the angle of the $r_e/\tilde{r}_e$-curve decreases, and eventually this curve coincides with the $(S_0 A)$ line, which corresponds to the saddle-node bifurcation in the time-averaged forced system (the stable fixed point $\tilde{S}_0$ merges with the unstable fixed point $\tilde{S}_1$ at the point $A$). With the further increase of excitability, the fixed points $\tilde{S}_0$ and $\tilde{S}_1$ disappear, and the dynamics of the forced system diverges.

### B. Numerical simulations of the low-dimensional system

To confirm the predictions of the geometrical analysis, we performed numerical simulation of the system described by (16). Simulation was performed during $T = 10\,s$ with the time step $\Delta t = 0.1$ ms. Periodic external forcing was turned on at $t_{osc} = 200$ ms.

Results of the simulation are presented in the Figure 2; top panel represents the firing rate traces, and the bottom panel – the membrane voltage traces (in both cases, the real part is shown). Red traces correspond to the excitatory population and the blue traces – to the inhibitory population. Thin solid lines represent the simulated dynamics and demonstrate forced oscillations for $t > t_{osc}$. Thick solid lines represent the simulated traces smoothed with 250-ms time window. Thick dashed lines represent predictions obtained from the geometrical analysis. We can see that the averaged simulated traces approach the values that are very close to the predicted ones.

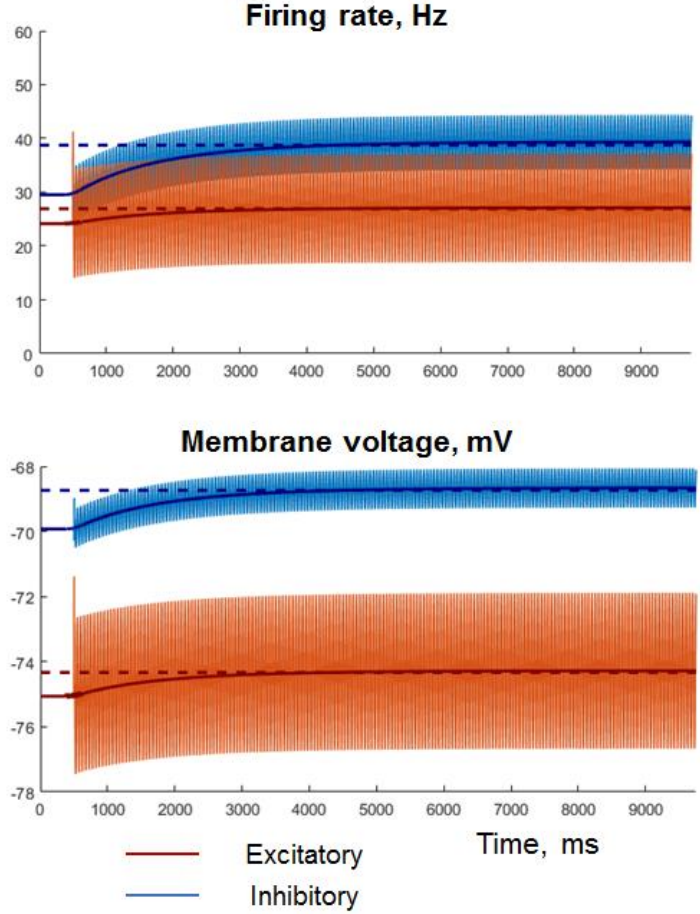

FIG 2. Results of simulation of Model 1. Red: excitatory population; blue: inhibitory population. Thin solid lines - simulated traces; thick solid lines – simulated traces averaged over 250 ms; thick dashed lines – predictions of the graphical analysis.

### C. Dependence of the steady-state shift on parameters

In this section, we analytically describe how certain parameters of the system affect the forced-induced shift of the time-averaged equilibrium.

Let us rewrite the system (45) in the following form:

$$\begin{cases} \tilde{I}_{NMDAe} = (\tilde{r}_e - P_{re})/Q_{re}^e \equiv p(\tilde{r}_e, \bar{\theta}) \\ \tilde{I}_{NMDAe} = I_{NMDAe}^{ss}(\tilde{r}_e, \bar{V}_e(\tilde{r}_e)) + D_e(\tilde{r}_e, \bar{V}_e(\tilde{r}_e)) \equiv q(\tilde{r}_e, \bar{\theta}) \end{cases}, \quad (46)$$

where $\bar{\theta} = \{\theta_i\}$ is a vector of parameters that define the functions $p$ and $q$. The steady-state $\tilde{r}_{e0}$ of the time-averaged forced system is given by the equation:

$$p(\tilde{r}_{e0}, \bar{\theta}) = q(\tilde{r}_{e0}, \bar{\theta}). \quad (47)$$

If a parameter $\theta_i$ is such that $q$ depends on it, but $p$ does not, then the dependence of $\tilde{r}_{e0}$ on $\theta_i$ could be obtained from the following expression:

$$\frac{\partial \tilde{r}_{e0}}{\partial \theta_i} = \frac{\partial q}{\partial \theta_i} \bigg/ \left( \frac{\partial p}{\partial \tilde{r}_{e0}} - \frac{\partial q}{\partial \tilde{r}_{e0}} \right). \quad (48)$$

As we mentioned in the previous section, if $\tilde{r}_{e0}$ is stable (which is the case of interest), then the term in the denominator of the right-hand part of (48) is positive. Consequently, as $\theta_i$ grows, sign of change of $\tilde{r}_{e0}$ and $q$ is the same.

Using (2), (32), and (34), we can express $q$ as follows:

$$q = J_{ee}^{NMDA} \left[ g_{NMDA}\left(\bar{V}_e(\tilde{r}_e)\right)\tilde{r}_e + \frac{1}{2}\left|r_e^A\right|^2 \alpha_e g'_{NMDA}\left(\bar{V}_e(\tilde{r}_e)\right)\cos\varphi_e + \frac{1}{4}\left|r_e^A\right|^2 \alpha_e^2 g''_{NMDA}\left(\bar{V}_e(\tilde{r}_e)\right)\tilde{r}_e \right],$$
(49)

where $\alpha_e = \left[c_{Va}\left(1+\omega^2\tau_{ra}^2\right)\right]/\left[c_{ra}\left(1+\omega^2\tau_{Va}^2\right)\right]$.

One can see that $q$ grows with $J_{ee}^{NMDA}$ and $\left|r_e^A\right|$. As $p$ does not depend on these parameters, we can conclude that $\tilde{r}_{e0}$ also grows as $J_{ee}^{NMDA}$ and $\left|r_e^A\right|$ increase.

On the other hand, if $p$ depends on a parameter $\theta_i$, but $q$ does not, the the dependence of $\tilde{r}_{e0}$ on $\theta_i$ could be obtained from the expression:

$$\frac{\partial \tilde{r}_{e0}}{\partial \theta_i} = -\frac{\partial p}{\partial \theta_i} \bigg/ \left( \frac{\partial p}{\partial \tilde{r}_{e0}} - \frac{\partial q}{\partial \tilde{r}_{e0}} \right).$$
(50)

Consequently, $\tilde{r}_{e0}$ grows with $\theta_i$ as $p$ decreases with $\theta_i$, and vice versa.

Using (12), we can calculate the derivatives of $p$ by the fast synaptic weights:

$$\begin{cases} \dfrac{\partial p}{\partial J_{ee}^{AMPA}} = -(\tilde{r}_{e0} - r_{e0})g_{me} \\[6pt] \dfrac{\partial p}{\partial J_{ie}^{AMPA}} = \dfrac{(\tilde{r}_{e0} - r_{e0})g_{me}c_{ri}J_{ei}}{(1+c_{ri}J_{ii})} \\[6pt] \dfrac{\partial p}{\partial J_{ei}} = \dfrac{(\tilde{r}_{e0} - r_{e0})g_{me}c_{ri}J_{ie}^{AMPA}}{(1+c_{ri}J_{ii})} \\[6pt] \dfrac{\partial p}{\partial J_{ii}} = -\dfrac{(\tilde{r}_{e0} - r_{e0})g_{me}c_{ri}^2 J_{ei}J_{ie}^{AMPA}}{(1+c_{ri}J_{ii})^2} \end{cases}.$$
(51)

Using the fact that $\tilde{r}_{e0} > r_{e0}$ when both $\tilde{r}_{e0}$ and $r_{e0}$ are stable, be can conclude from (51) that $\tilde{r}_{e0}$ increases with $J_{ee}^{AMPA}, J_{ii}$ (i.e. when self-excitation in the system gets stronger), and decreases with $J_{ie}^{AMPA}, J_{ei}$ (i.e. when self-inhibition gets stronger).

As we have seen, the requirement of stability of the time-averaged forced system strongly limits the possible shift of the excitatory firing rate produced by the external forcing. However, the shift of the inhibitory firing rate could be made sufficiently large. Using (11) for $I_{NMDAe} = I_{NMDAe0}$ and $I_{NMDAe} = \tilde{I}_{NMDAe0}$, we can express the firing rate shifts as follows:

$$\begin{cases} \tilde{r}_{e0} - r_{e0} = Q_{re}^e \left( \tilde{I}_{NMDAe0} - I_{NMDAe0} \right) \\ \tilde{r}_{i0} - r_{i0} = Q_{ri}^e \left( \tilde{I}_{NMDAe0} - I_{NMDAe0} \right) \end{cases}.$$
(52)

Consequently, the inhibitory firing rate shift could be derived from the excitatory firing rate shift:

$$\tilde{r}_{i0} - r_{i0} = \frac{Q_{ri}^e}{Q_{re}^e}(\tilde{r}_{e0} - r_{e0}).$$
(53)

If we increase $\tilde{J}_{ie}^{AMPA}$ and decrease $\tilde{J}_{ei}$ is such way that the product $\tilde{J}_{ie}^{AMPA}\tilde{J}_{ei}$ does not change (and also adjust the external inputs in order to keep $u_{e0}, u_{i0}, \sigma_{e0}^2, \sigma_{i0}^2, r_e^A, r_i^A$ constant, as it was described above), then $Q_{ri}^e$ will increase, while $Q_{re}^e$ and $\tilde{r}_{e0}$ will stay the same. Consequently, by increasing $\tilde{J}_{ie}^{AMPA}$ (with the corresponding adjustment of other parameters), we can make the deviation $\tilde{r}_{i0} - r_{i0}$ as large as we want.

## V. ANALYSIS OF SYSTEM WITH NMDA SYNAPSES ON THE EXCITATORY AND INHIBITORY POPULATIONS (MODEL 2)

In this section, we analyze the Model 2 that contains NMDA receptors both on the excitatory and inhibitory neurons. First, we note that the excitatory-to-inhibitory (E-I) NMDA coupling $\tilde{J}_{ie}^{NMDA}$ is much weaker in the model than the excitatory-to-excitatory (E-E) NMDA coupling $\tilde{J}_{ee}^{NMDA}$, so the main effect of the external forcing on the system is still excitatory. The role of the E-I NMDA coupling here is to make the system more robust, which allows the fast subsystem to be more excitable without having divergent dynamics in the presence of the external forcing. In the Model 2, the fast subsystem was made more excitable compared to the Model 1 by decreasing the E-I AMPA coupling $\tilde{J}_{ie}^{AMPA}$.

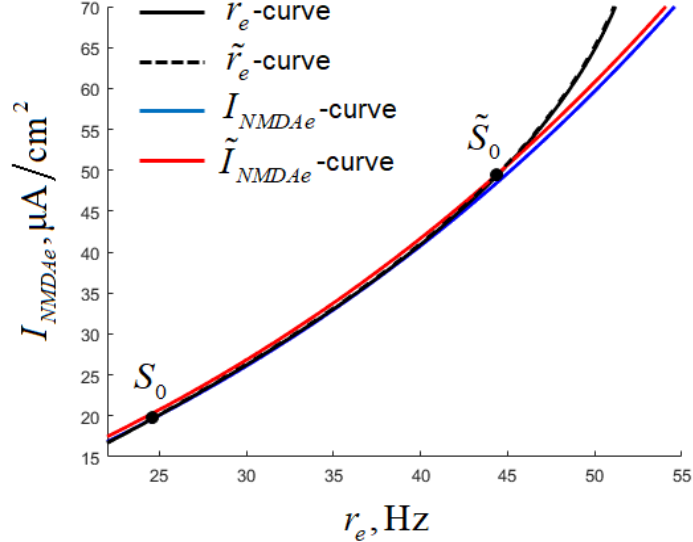

FIG 3. Part of the phase plane with $r_e, \tilde{r}_e, I_{NMDAe}, \tilde{I}_{NMDAe}$-curves for the Model 2. Solid black line - $r_e$-curve; dashed black line - $\tilde{r}_e$-curve; solid blue line - $I_{NMDAe}$-curve; solid red line - $\tilde{I}_{NMDAe}$-curve; $S_0, \tilde{S}_0$ - equilibria of the unforced and the time-averaged forced systems, respectively.

### A. Phase-plane analysis

The phase plane for the Model 2 is presented in the Figure 3. The legend is the same as in the Figure 1A (which represents the same picture for the Model 1), with the difference that the $\tilde{r}_e$-curve now differs from the $r_e$-curve; the $\tilde{r}_e$-curve is represented in the Figure 3 by the dashed black line, and the $r_e$-curve is represented by the solid black line.

Comparing the Figures 1 and 3, one can see two major differences between the Model 1 and the Model 2. First, as it was demonstrated above, the $\tilde{r}_e$-curve for the Model 1 is a straight line, while this curve for the Model 2 is concave. Consequently, the position of the time-averaged forced equilibrium $\tilde{S}_0$ in the Model 1 is limited by the point, denoted in the Figure 1B as $A$ (which is the point on the $\tilde{I}_{NMDAe}$-curve such that the line $(S_0 A)$ is tangent to the $\tilde{I}_{NMDAe}$-curve); at the same time, there is no such limitation in the Model 2.

The second difference is that in the Model 1 the unstable fixed point $\tilde{S}_1$ of the time-averaged forced system exists near the equilibrium point $\tilde{S}_0$, while in the Model 2 the $\tilde{r}_e$-curve quickly goes away from the $\tilde{I}_{NMDAe}$-curve in the region above $\tilde{S}_0$, so there is no unstable point $\tilde{S}_1$ near $\tilde{S}_0$. Thus, we expect that no saddle-fold bifurcation (and, consequently, no divergent dynamics) will occur if we make the fast subsystem more excitable (as we observed in the Model 1). In the following section, we demonstrate that it is indeed the case, and that the system loses its stability via Hopf bifurcation, and starts to generate very slow oscillations.

We should note that stability of $S_0$ and $\tilde{S}_0$ for the case of $\tilde{J}_{ie}^{NMDA} \neq 0$ does not follow from simple geometrical considerations, so it should be checked explicitly by numerically calculating the eigenvalues of the matrices given by the expressions (39) and (41). We did it for the Model 2, and confirmed that both $S_0$ and $\tilde{S}_0$ are stable.

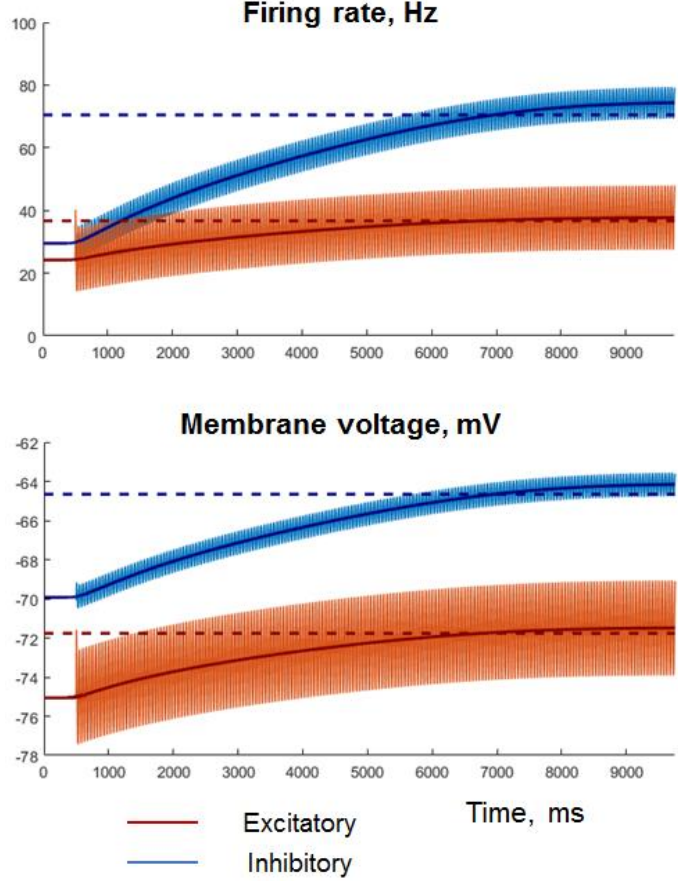

FIG 4. Results of simulation of Model 2. Red: excitatory population; blue: inhibitory population. Thin solid lines - simulated traces; thick solid lines – simulated traces averaged over 250 ms; thick dashed lines – predictions of the graphical analysis.

### B. Numerical simulations of the low-dimensional system

We confirmed the predictions of the geometrical analysis for the Model 2 by numerical simulation of the system (16), using the same simulation parameters as we did for the simulation of the Model 1. The results of simulation are represented in the Figure 4; the legend is the same as in the Figure 2.

One can see that the simulation results are in the good agreement with the predictions; however the prediction error is a bit larger than for the Model 1. This error is caused by the fact that the prediction is obtained using time scale separation which assumes the infinite ratio between the slow and the fast time scales, while in our system this ratio is, obviously, finite. To prove that the error is caused by the finite time scale ratio, we increased $\tau_{NMDA}$ (making it eight times larger) and proportionally decreased $\tilde{J}_{ee}^{NMDA}$, $\tilde{J}_{ie}^{NMDA}$ (making them eight times smaller), so $J_{ee}^{NMDA}$, $J_{ie}^{NMDA}$ did not change (see (5)), and the predicted shift was the same as for the Model 2. Simulation of the modified model (see Appendix) demonstrated almost perfect match with the prediction.

### C. Dependence of the steady-state shift on parameters

In this section we numerically explore dependence of the forced-induced shift of the time-averaged equilibrium $\Delta r_{e0} = \tilde{r}_{e0} - r_{e0}$ on certain parameters of the system having NMDA receptors both on the excitatory and the inhibitory population.

The dependence of $\Delta r_{e0}$ on the slow synaptic weights $J_{ee}^{NMDA}, J_{ie}^{NMDA}$ is presented in the Fig. 5(a), on the external forcing amplitudes $r_e^A, r_i^A$ – in the Fig. 5 (b), on the fast excitatory synaptic weights $J_{ee}^{AMPA}, J_{ie}^{AMPA}$ – in the Fig. 6(a), on the fast inhibitory synaptic weights – in the Fig. 6(b). The borders of instability of the unforced slow subsystem are denoted by the black lines; of the unforced fast subsystem – by the green lines; of the time-averaged forced system – by the red lines.

Fig. 5,6 suggest that $\Delta r_{e0}$ increases with $J_{ee}^{NMDA}, J_{ee}^{AMPA}, J_{ii}$ (i.e. when there is more excitation in the system) and decreases with $J_{ie}^{NMDA}, J_{ie}^{AMPA}, J_{ei}$ (i.e. when there is more inhibition in the system). Also, $\Delta r_{e0}$ increases with $r_e^A$ (i.e. when the excitatory population is more activated by the external forcing) and decreases with $r_i^A$ (i.e. when the inhibitory population is more activated by the external forcing).

It is seen that the maximal $\Delta r_{e0}$ is always observed near the borders of instability of the slow subsystem. The time-averaged forced system becomes unstable as $J_{ee}^{NMDA}, J_{ee}^{AMPA}, J_{ii}, r_e^A$ increase and $J_{ie}^{NMDA}, J_{ie}^{AMPA}, J_{ei}, r_i^A$ decrease. The unforced slow subsystem becomes unstable as $J_{ee}^{NMDA}, J_{ee}^{AMPA}, J_{ii}$ increase and $J_{ie}^{NMDA}, J_{ie}^{AMPA}, J_{ei}$ decrease. The unforced fast subsystem becomes unstable for very large values of $J_{ee}^{AMPA}$ and for combinations of very moderately large values of $J_{ee}^{AMPA}$ and very small values of $J_{ie}^{AMPA}$. Interestingly, there is an optimal combination of $J_{ee}^{NMDA}, J_{ie}^{NMDA}$, for which the maximal $\Delta r_{e0}$ is reached [see Fig. 5 (a)].

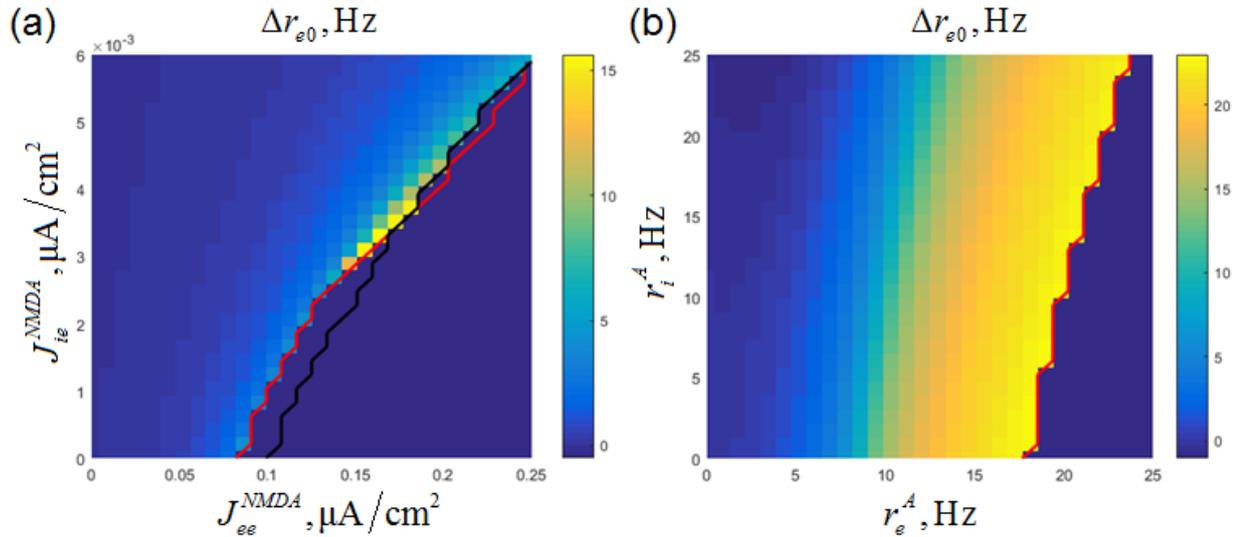

FIG 5. Dependence of the forcing-induced shift of the time-averaged equilibrium $\Delta r_{e0}$ on the following parameters: (a) the slow synaptic weights $J_{ee}^{NMDA}, J_{ie}^{NMDA}$, (b) the external forcing amplitudes $r_e^A, r_i^A$. Black line: the border of stability of the unforced slow subsystem; red line: the border of stability of the time-averaged forced system. The instability region is filled by the dark-blue color.

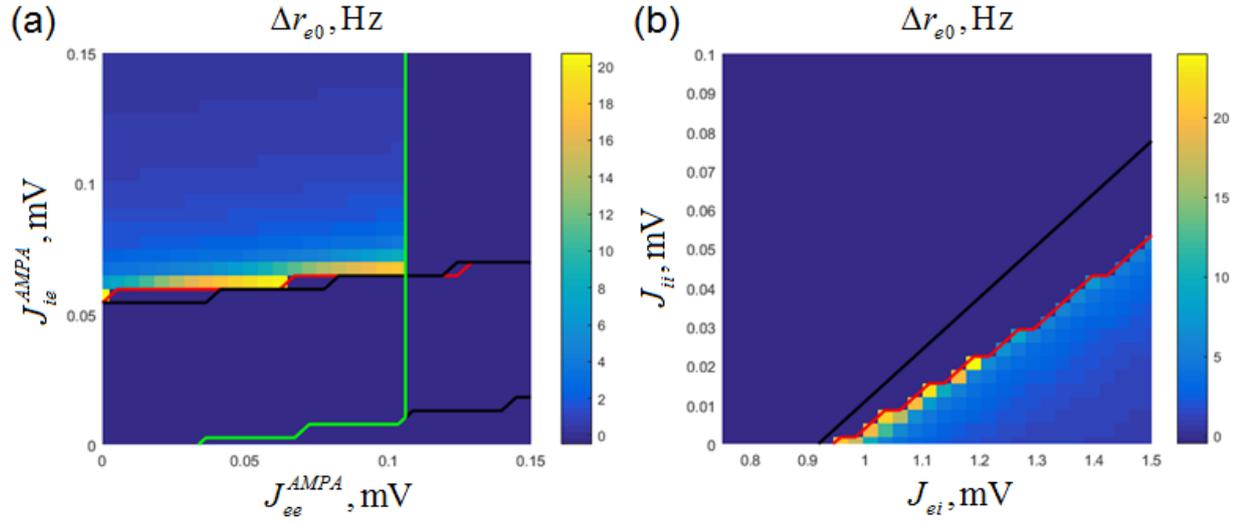

FIG 6. Dependence of the forcing-induced shift of the time-averaged equilibrium $\Delta r_{e0}$ on the following parameters: (a) the fast excitatory synaptic weights $J_{ee}^{AMPA}, J_{ie}^{AMPA}$, (b) the fast inhibitory synaptic weights $J_{ei}, J_{ii}$. Black line: the border of stability of the unforced slow subsystem; red line: the border of stability of the time-averaged forced system; green line: the border of stability of the unforced fast subsystem. The instability region is filled by the dark-blue color.

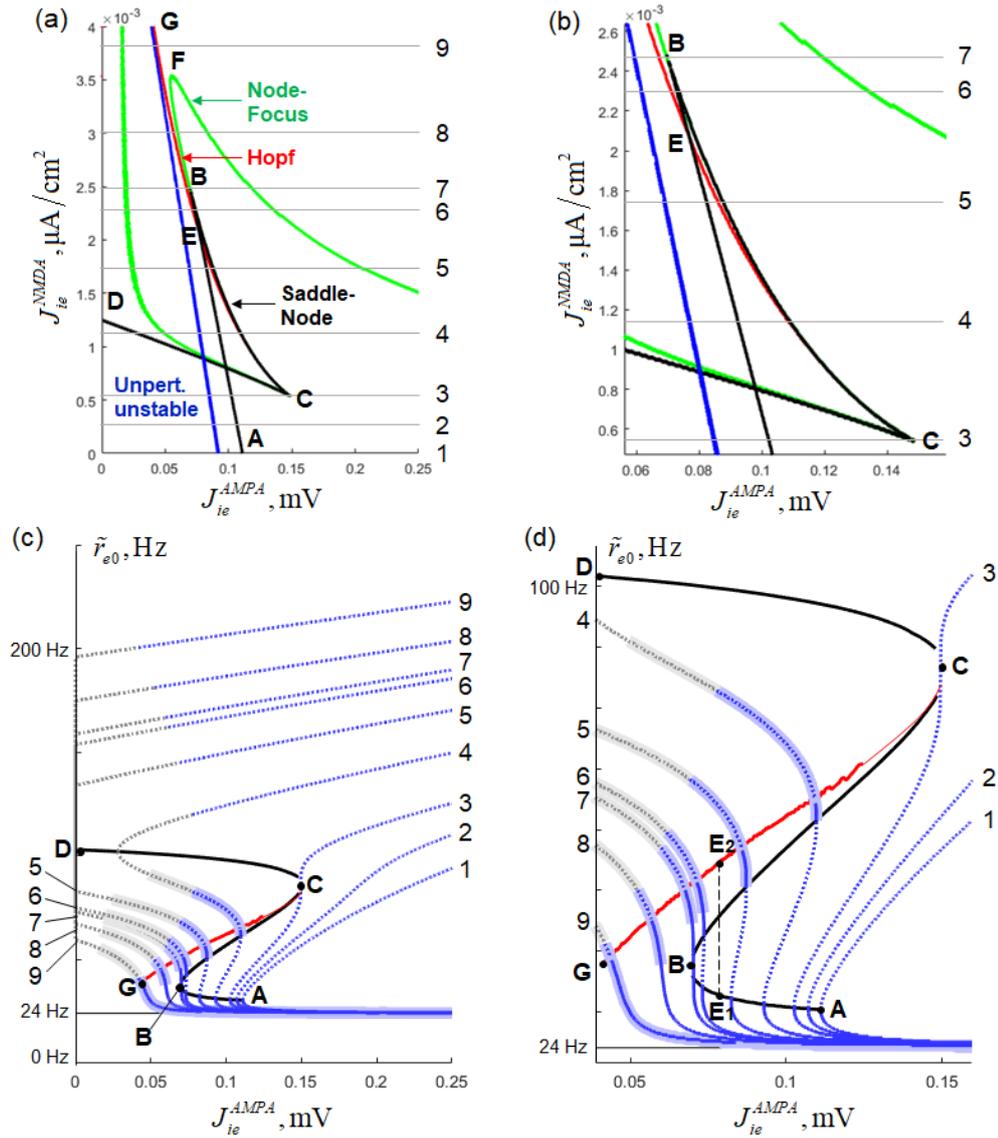

FIG. 7. (a,b) Bifurcation diagram of the time-averaged forced system. Horizontal axis: $J_{ei}^{AMPA}$, vertical axis: $J_{ie}^{NMDA}$. Black line ABCD: saddle-node bifurcations; B and C are the cusp points. Red line CG: Hopf bifurcations. Green line: Focus-to-node transitions. The point E is the intersection between CG and AB. In the region to the left of the blue line, the unforced system is unstable. Horizontal lines: the probed values of $J_{ie}^{NMDA}$; they are denoted by the numbers 1 – 9 at the right parts of the panels. (c,d) Blue / grey lines: one-dimensional (1-D) bifurcation diagrams (dependence of the time-averaged forced $\tilde{r}_e$ equilibria on $J_{ei}^{AMPA}$) for various values of $J_{ie}^{NMDA}$. Diagrams are marked by the numbers that correspond to horizontal lines in the Fig. 7(a). Solid lines: an equilibrium is stable; dashed lines: unstable. Blue color: the unforced equilibrium is stable; grey color: unstable. Superimposed thick lines: an equilibrium is a focus. Red line CG connects all Hopf bifurcation points; black line ABCD connects all saddle-node points; the meaning of A,B,C,D,G is the same as in the Fig. 7(a). $E_1$, $E_2$ – points on the Ab and CG lines, respectively, at $J_{ei}^{AMPA}$ value that corresponds to the point E in the Fig. 7(a). The numbers 1 – 9 in the left and the right parts of the panels correspond to the numbers in the Fig. 7(a,b) and link the curves with the selected $J_{ie}^{NMDA}$ levels.

Now we describe the dependence of the system's behavior on the slow E-I NMDA coupling weight $\tilde{J}_{ie}^{NMDA}$ and on the excitability of the fast subsystem (varied via the fast E-I AMPA coupling weight $\tilde{J}_{ie}^{AMPA}$). We analyzed the bifurcations that occur in the time-averaged forced system when $\tilde{J}_{ie}^{NMDA}$ and $\tilde{J}_{ie}^{AMPA}$ are varied and explored the consequences of these bifurcations for the ability to excite the system by external periodic forcing without destabilizing it. The main purpose of this analysis is to define how strongly the equilibrium could be shifted by the external forcing without loss of stability under various parameter combinations.

In the Fig. 7(a), the two-dimensional (2-D) bifurcation diagram for the time-averaged forced system is presented, with $\tilde{J}_{ie}^{AMPA}$ on the horizontal axis and $\tilde{J}_{ie}^{NMDA}$ on the vertical axis. In the Fig. 7(b), a region of the diagram is zoomed in. The region to the left of the blue line in the Fig. 7(a,b) corresponds to instability of the unforced equilibrium $r_{e0}$, and it is irrelevant for our analysis. The black line corresponds to saddle-node bifurcations: two fixed points merge and disappear as the line AB or BC is crossed from the right to the left, and as the line BC is crossed from the left to the right. In the region above the green line in the Fig. 7(a,b), one of the fixed points is a focus, so the system demonstrates either damped or sustained slow oscillations. This focus is stable to the right of the red line CG and unstable to the left of it; the red line CG itself corresponds to the supercritical Hopf bifurcations. The Intersection of the AB and CG lines is denoted by E.

To provide deeper understanding of the system's behavior, we fixed $\tilde{J}_{ie}^{NMDA}$ at various levels and analyzed how the fixed points of the system depend on $\tilde{J}_{ie}^{AMPA}$. The values of $\tilde{J}_{ie}^{NMDA}$ we used are shown in the Fig. 7(a,b) by horizontal lines and marked by numbers. The corresponding one-dimensional (1-D) bifurcation diagrams are shown in the Fig. 7(c) by the blue / grey lines [and marked by the same numbers as in the Fig. 7(a,b)], with $\tilde{J}_{ie}^{AMPA}$ on the horizontal axis, and the steady-state values of $\tilde{r}_e$ on the vertical axis. A region of interest of the Figure 5(c) is shown in more detail in the Fig. 7(d). Stable parts of the diagrams in the Fig. 7(c,d) are represented by solid lines, and the unstable parts – by dashed lines. Superimposed transparent thick lines denote parts of the diagrams that correspond either to stable or unstable foci. A diagram parts have grey color if they are located in the range of $\tilde{J}_{ie}^{AMPA}$ values for which the unforced system is unstable. All points of saddle-node bifurcations are connected by the black line, and all points of Hopf bifurcations – by the red line; the points A,B,C,D,G have the same meaning as in the Fig. 7(a,b). Please note that point E in the Fig. 7(a,b) is not a codimension 2 bifurcation, because the lines AB and CG correspond to bifurcations that occur on the different branches of 1-D diagrams. In the Fig. 7(d), the bifurcation points on the two stable branches for the $\tilde{J}_{ie}^{AMPA}$ value that corresponds to the point E, are denoted by $E_1$ and $E_2$.

For all $\tilde{J}_{ie}^{NMDA}$ values below the cusp point B [see lines 1 – 6 in the Fig. 7(a)], the corresponding 1-D diagrams contain a stable branch that terminates by a saddle-node bifurcation [see the lowest parts of the diagrams 1 – 6 in the Fig. 7(c,d) that are located below the black line AB]. For $\tilde{J}_{ie}^{NMDA}$ values below the cusp point C [lines 1 and 2 in the Fig. 7(a)], this is the only stable branch of the 1-D diagram. From the Fig. 7(c,d), one can see that the lower branches go up as $\tilde{J}_{ie}^{AMPA}$ decreases. Consequently, the maximal values of $\tilde{r}_{e0}$ for the diagrams 1 and 2 are achieved at the $\tilde{J}_{ie}^{AMPA}$ values that correspond to saddle-node bifurcation; these maximal $\tilde{r}_{e0}$ values lie on the black line AB.

For $\tilde{J}_{ie}^{NMDA}$ values between the cusp points C and B [lines 4 – 6 in the Fig. 7(a)], the corresponding 1-D diagrams still contain the aforementioned lower stable branch, but, in addition to that, a part of another branch with higher $\tilde{r}_e$ is also stable; in the Fig. 7(c,d), these stable branch parts are located between the black line BC and the red line CG. Bistability occurs for combinations of $\tilde{J}_{ie}^{NMDA}$ and $\tilde{J}_{ie}^{AMPA}$ that lie in the region CEB in the Fig. 7(a,b). In principle, forced oscillations may cause the system settle onto the upper stable branch, but our numerical simulations demonstrated that in most cases the lower branch is selected (however, a counter-example is presented in the Appendix). Furthermore, the bistability region CEB is very narrow, so, even if selection of the upper branch is possible, it requires extremely fine tuning of parameters.

For $\tilde{J}_{ie}^{NMDA}$ values below the point E [lines 4 and 5 in the Fig. 7(a)], the Hopf bifurcation line CG (red) in the Fig. 7(a,b) lies to the right of the saddle-node bifurcation line AB (black). Accordingly, as $\tilde{J}_{ie}^{AMPA}$ decreases, the upper stable branch of the 1-D diagram [see diagrams 4 and 5 in the Fig. 7(c,d)] loses its stability (crossing the red line CG) before the lower stable branch disappears (crossing the black line

AB). Consequently, if we assume that the upper branch is never selected under bistable conditions, than the maximal values of $\tilde{r}_{e0}$ that could be reached by varying $\tilde{J}_{ie}^{AMPA}$ are, again, located on the AB line.

For $\tilde{J}_{ie}^{NMDA}$ values between the points E and B [line 6 in the Figure 5(a)], the Hopf line CG goes to the left of the saddle-node line AB. In this case, as $\tilde{J}_{ie}^{AMPA}$ decreases, the lower stable branch of the 1-D diagram [see diagram 6 in the Fig. 7(c,d)] disappears (crossing the black line AB), but for even smaller $\tilde{J}_{ie}^{AMPA}$ values, the upper stable branch still exists, until it loses stability via Hopf bifurcation (crossing the red line CG). Consequently, the maximal values of $\tilde{r}_{e0}$ that could be achieved by varying $\tilde{J}_{ie}^{AMPA}$, are located on the CG line.

For $\tilde{J}_{ie}^{NMDA}$ values above the cusp point C [lines 8 and 9 in the Fig. 7(a)], only one stable branch of the 1-D diagram exists [see diagrams 8 and 9 in the Fig. 7(c,d)]; this branch loses its stability via Hopf bifurcation as $\tilde{J}_{ie}^{AMPA}$ decreases (crossing the red line CG). The maximal values of $\tilde{r}_{e0}$ that could be achieved by varying $\tilde{J}_{ie}^{AMPA}$, are, again, located on the CG line.

In summary, we can distinguish two regions of $\tilde{J}_{ie}^{NMDA}$ values, based on the location of the maximal $\tilde{r}_{e0}$ that could be achieved by varying $\tilde{J}_{ie}^{AMPA}$. First, for a $\tilde{J}_{ie}^{NMDA}$ value below the point E in the Fig. 7(a,b), the maximal $\tilde{r}_{e0}$ is achieved at the value of $\tilde{J}_{ie}^{AMPA}$ that corresponds to the saddle-node bifurcation that terminates the lower stable branch of the 1-D diagram at its left side [see diagrams 1 – 5 in the Fig. 7(c,d)]. All such $\tilde{r}_{e0}$ values are located on the curve $AE_1$ in the Fig. 7(d). Second, for a $\tilde{J}_{ie}^{NMDA}$ value above the point E in the Fig. 7(a,b), the maximal $\tilde{r}_{e0}$ is achieved at the value of $\tilde{J}_{ie}^{AMPA}$ that corresponds to the Hopf bifurcation [see diagrams 6,7 in the Fig. 7(c,d)]. All such $\tilde{r}_{e0}$ values are located on the red line $E_2G$ in the Fig. 7(d). Thus, the dashed line $AE_1E_2G$ in the Fig. 7(d) defines maximal $\tilde{r}_{e0}$ values that could be, in principle, achieved by the external forcing without destabilizing the system. As one can see, the largest of these values is reached at the point $E_2$; this value corresponds to the firing rate shift (relatively to the unforced equilibrium) that is equal to approximately 20 Hz.

## VI. CONCLUSION

Numerous theoretical concepts of neural processing (e.g., selectivity to certain information) are formulated in terms of average firing rates. At the same time, brain activity demonstrates collective oscillatory patterns that correlate with functional states, task requirements and behavioral features. In order to build a theory that reconciles rate-based neural coding with functional role of oscillations in computations and information routing, one should consider a non-linear mechanism that converts oscillatory power into tonic firing rate shifts. In this paper, we explored a potential role that NMDA-synapses with non-linear behavior could play in shifting of the average level of population activity in the presence of external oscillatory input. We considered an excitatory-inhibitory population model with the population firing rates, mean membrane voltages, and NMDA-currents as the dynamical variables. NMDA-current depended both on presynaptic firing rate and postsynaptic voltage. In order to delineate the NMDA-related effects from the effects of non-linearity of spike generation mechanism, we linearized the dependences of the firing rate and the mean voltage on the synaptic input about the unperturbed equilibrium of the system.

We applied the time scale separation and the time-averaging method, which allowed us to analytically derive expressions for the mean firing rate shift, induced by the oscillatory input to the system. Under realistic model parameters, we found that such shift is mostly produced by joint effect of the presynaptic firing rate and the postsynaptic voltage oscillations occurring almost in phase, and not by the non-linear dependence of NMDA-current on the postsynaptic voltage. Our analytical predictions were confirmed by direct numerical simulations.

We considered two models. In the Model 1, NMDA receptors were located exclusively on the excitatory neurons, while in the Model 2, they were located both on the excitatory and the inhibitory neurons. Using phase plane analysis based on our analytical results, we geometrically proved that the oscillation-induced firing rate shift is strongly limited by stability requirements in the Model 1. Adding even very weak NMDA input to the inhibitory population allows to overcome this limitation and achieve pronounced firing rate shift (up to 20 Hz) without destabilizing the system. Finally, we explored the parameter space and found optimal regions of parameters (NMDA input weights to both populations, AMPA input weight to the inhibitory population), in which the strongest firing rate could be achieved under the same amplitude of the entrained oscillations.

It is clear that many neural processes could potentially link oscillatory activity to mean firing rate modulations. As the examples, one could propose non-linearity of spike generation, non-linear behavior of various slow voltage-dependent ion channels, transition between spiking and bursting, or synaptic plasticity. Furthermore, the character of such link should critically depend on microconnectivity patterns. Nevertheless, the NMDA-based mechanism that we proposed here is rather general and should play certain role in oscillatory-induced firing rate shifts alongside with other potential mechanisms in most neural configurations. However the extent to which the NMDA-related non-linearity is involved in these shifts is a subject of future experimental research.

## APPENDIX A: PARAMETER DERIVATION

### 1. Linearization coefficients and the CV

In our numerical computations, we assumed that the source of the input variance is mainly external, so this variance does not depend on the state of the system. Following this assumption, we simulated a single excitatory (and a single inhibitory) leaky integrate-and-fire neuron that received Gaussian white noise with the amplitude $\sigma_{a0}^2$ and the tonic input $u_a$. We probed several values of $u_a$ around $u_{a0}$. The simulations were governed by the following equations:

$$\begin{cases} \tau_{ma} \dfrac{dv_a}{dt} = -v_a + E_{La} + u_a + \sigma_{a0}\eta_a(t) \\ v_a > V_a^{th} : v_a \leftarrow V_a^r \end{cases}, \tag{A1}$$

where $v_a$ - the membrane voltage, $\tau_{ma}$ - the membrane time constant, $E_{La}$ - the resting potential, $\eta_a(t)$ - white noise with zero mean and unit standard deviation, $V_a^{th}$ - the spiking threshold, $V_a^r$ - the reset voltage.

For each value of $u_a$, we calculated the resulting mean membrane voltage $\hat{V}_a^{ss,0}(u_a)$ and the firing rate $\hat{r}_a^{ss,0}(u_a)$. Given these empirical dependencies, we numerically estimated their derivatives at $u_{a0}$, thus determining the constants $c_{ra}$ and $c_{Va}$.

Also, from the simulated spike train for $u_a = u_{a0}$ we calculated the coefficient of inter-spike interval variation $CV_{a0}$ which characterizes irregularity of the spike train.

### 2. Population time constants

In order to find population time constants $\tau_{ra}$ and $\tau_{Va}$, we considered $N = 5000$ uncoupled neurons whose dynamics are governed by (A1) with $u_a = u_{a0}$. We repeated this simulation 20 times. For each time bin, we averaged the membrane voltages over neurons and trials, thus computing the empirical temporal dynamics of the population membrane voltage $\hat{V}_a(t)$. We also averaged the numbers of spikes produced by the network over trials, thus computing the empirical temporal dynamics of the population firing rate $\hat{r}_a(t)$. After $\hat{r}_a(t)$ and $\hat{V}_a(t)$ stabilized (given $u_a = u_{a0}$), we increased $u_a$ by $\Delta u = 2$ mV, continued the simulations, and calculated $\hat{r}_a(t)$ and $\hat{V}_a(t)$ for the subsequent time moments, until they stabilized again. Then we fitted the transitions of $\hat{r}_a(t)$ and $\hat{V}_a(t)$ produced by the increase of $u_a$ by exponential functions, from which we got $\tau_{ra}$ and $\tau_{Va}$, respectively. The resulted $\hat{r}_a(t)$ and $\hat{V}_a(t)$ for excitatory and inhibitory network (and the corresponding exponential fits) are presented in the Figures 8(a) and 8(b), respectively.

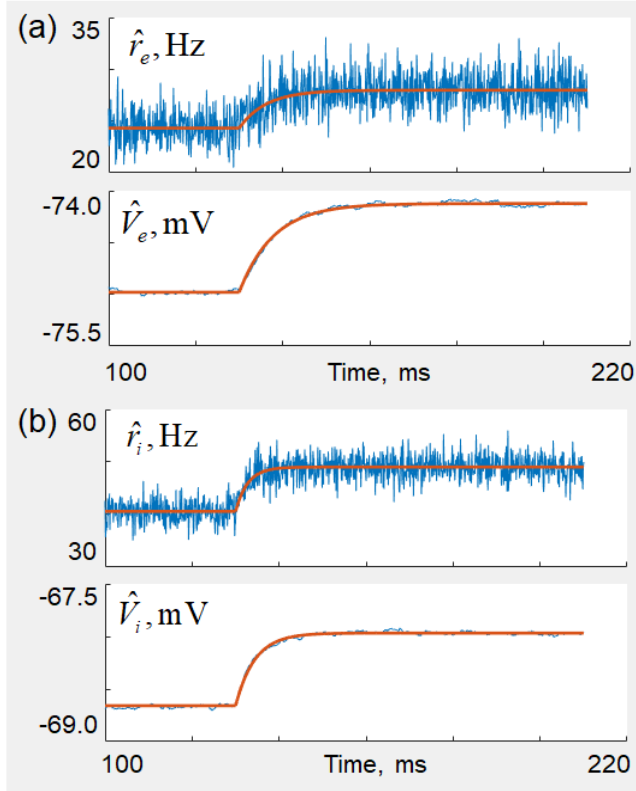

FIG. 8. Results of simulation of an uncoupled population of excitatory (a) and inhibitory (b) neurons. Top panels: dynamics of the population firing rate, bottom panels: dynamics of the mean membrane voltage. Blue lines: simulation results; red lines: exponential fits.

### 3. Parameters of the external inputs

Using the single-neuron simulation (as in the previous paragraphs), we numerically defined the steady-state population firing rate $r_{a0}$ and the population membrane voltage $V_{a0}$ given the input $u_a = u_{a0}$:

$$\begin{cases} r_{a0} = \hat{r}_a^{ss,0}(u_{a0}) \\ V_{a0} = \hat{V}_a^{ss,0}(u_{a0}) \end{cases}. \tag{A2}$$

From the found values of $r_{a0}$ and $V_{a0}$, using (2), we calculated the steady-state NMDA-current $I_{NMDAa0}$:

$$I_{NMDAa0} = I_{NMDAa}^{ss}(r_{a0}, V_{a0}). \tag{A3}$$

Then we determined mean total inputs $h_e, h_i$, mean external inputs $h_{xe}, h_{xi}$, and external input variances $\sigma_{xe}^2, \sigma_{xi}^2$:

$$\begin{cases} h_e = u_{e0} - \left( \tilde{J}_{ee} K_{ee} \tau_{me} r_e - \tilde{J}_{ei} K_{ei} \tau_{me} r_i + I_{NMDA,0}/g_{me} \right) \\ h_i = u_{i0} - \left( \tilde{J}_{ie} K_{ie} \tau_{mi} r_e - \tilde{J}_{ii} K_{ii} \tau_{mi} r_i \right) \\ h_{xe} = h_e - E_{Le} \\ h_{xi} = h_i - E_{Li} \\ \sigma_{xe}^2 = \sigma_{e0}^2 - \frac{1}{2} \left( \tilde{J}_{ee}^2 K_{ee} \tau_{me} r_e + \tilde{J}_{ei}^2 K_{ei} \tau_{me} r_{ii} \right) \\ \sigma_{xi}^2 = \sigma_{i0}^2 - \frac{1}{2} \left( \tilde{J}_{ie}^2 K_{ie} \tau_{mi} r_e + \tilde{J}_{ii}^2 K_{ii} \tau_{mi} r_{ii} \right) \end{cases}. \tag{A4}$$

Although $h_{xe}, h_{xi}, \sigma_{xe}^2, \sigma_{xi}^2$ do not participate directly in our analysis, we use them to check whether these values could be achieved under realistic conditions.

In order to calculate the external oscillatory inputs, we used the real-valued amplitudes $|r_e^A|, |r_i^A|$ of the forced oscillations delivered to the excitatory and inhibitory populations, respectively, as well as the phase lag $\gamma_r$ between them, and constructed the corresponding complex-valued amplitudes as follows:

$$\begin{cases} r_e^A = |r_e^A| \\ r_i^A = |r_i^A|\exp(i\psi_{ei}) \end{cases}. \tag{A5}$$

Then we calculated the complex amplitudes $h_e^A, h_i^A$ of the oscillatory external inputs using the expression (21).

### APPENDIX B: MODEL WITH VERY SLOW NMDA DYNAMICS

For the Model 2, we observed a deviation between the prediction and the simulation results (see Figure 4). In order to confirm that this deviation is due to finite ratio between the slow and the fast time scales, we considered a modified version of Model 2, in which the time constant of NMDA dynamics was made very large. The slow synaptic weights were appropriately rescaled, so the predicted effect of the external forcing was the same as for the Model 2. In summary, we following parameters differed from the Model 2:

$\tau_{NMDA} = 1600$ ms

$J_{ee}^{NMDA} = 0.01875$ μA/cm$^2$

$J_{ie}^{NMDA} = 0.000375$ μA/cm$^2$

Due to very slow dynamics, we simulated the modified model for a longer period of time $T = 120$ s.

The result is presented in the Figure 9. One can see that the prediction error is much smaller than in the Figure 4 (for the Model 2), so the prediction is almost perfect.

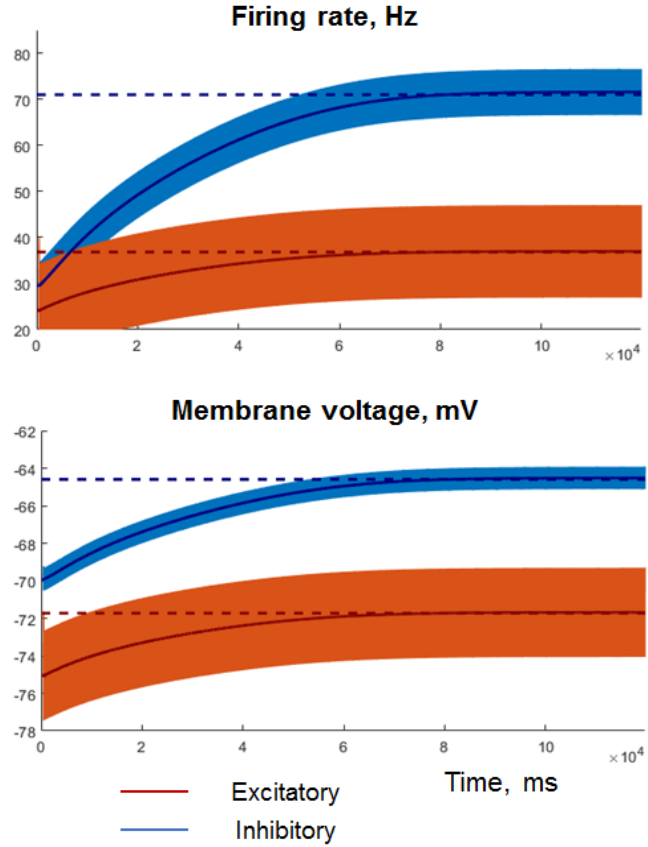

FIG. 9. Results of simulation of a model analogous to Model 2, but with very slow NMDA time constant.

Red: excitatory population; blue: inhibitory population. Thin solid lines - simulated traces; thick solid lines – simulated traces averaged over 250 ms; thick dashed lines – predictions of the graphical analysis.

## APPENDIX C: DESCRIPTION OF THE BIFURCATION DIAGRAMS

In this section, we provide complete description of the bifurcation diagrams depicted in the Fig. 7. We successively go from small to large values of $\tilde{J}_{ie}^{NMDA}$ [referred to by numbers from 1 to 9 in the Fig. 7(a)] and describe the events that occur as $\tilde{J}_{ie}^{AMPA}$ is varied.

The values 1 and 2 represent the case, in which only one saddle-node bifurcation is possible. In the Fig. 7(a), it corresponds to crossing of the line AB by lines 1 and 2 from the right to the left. In the Fig. 7(c,d), one can see that, as $\tilde{J}_{ie}^{AMPA}$ decreases, the stable branches of diagrams 1 and 2 (representing stable nodes) merge with the corresponding unstable branches (representing saddle points) and disappear. For even smaller values of $\tilde{J}_{ie}^{AMPA}$, the dynamics becomes divergent. The Model 1 considered in the paper, belongs to this case.

For the value 3, in addition to the aforementioned saddle-node bifurcation, one can observe the cusp which is denoted by the point C in the Fig. 7(a-d). For larger values of $\tilde{J}_{ie}^{NMDA}$, two saddle-node lines CB and CD grow from the cusp point C. In according to this, the diagram 4 in the Fig. 7(c,d) contains three saddle-node bifurcation points. Let us describe the sequence of events that occur as $\tilde{J}_{ie}^{AMPA}$ decreases from the maximal value to zero.

- For large $\tilde{J}_{ie}^{AMPA}$, two equilibria exist: the first one is the stable node that corresponds to the lower branch of the diagram 4 in the Fig. 7(c,d), and the second one is the saddle that corresponds to the upper branch of the diagram 4.

- As $\tilde{J}_{ie}^{AMPA}$ decreases, the line 4 in the Fig. 7(a) [as well as the diagram 4 in the Fig. 7(c,d)] crosses the black line BC. After this, two additional middle branches in the diagram 4 appear (as a result of a saddle-node bifurcation). The lower-middle branch corresponds to a saddle point. The upper-middle branch corresponds to a stable node just to the left of the bifurcation point, and very soon undergoes the subsequent changes.

- As $\tilde{J}_{ie}^{AMPA}$ further decreases, the line 4 in the Fig. 7(a) crosses the green line, and the fixed point corresponding to the upper-middle branch of the diagram 4 in the Fig. 7(c,d) transforms from the stable node to the stable focus (denoted by superimposed thick line).

- After this, the line 4 in the Fig. 7(a) crosses the red line CG, after which the fixed point that corresponds to the upper-middle branch of the diagram 4 in the Fig. 7(c,d) becomes an unstable focus via supercritical Hopf bifurcation (the branch is now denoted by dashed line with superimposed thick line). We should note that all three aforementioned events (the saddle-node bifurcation, the node-to-focus transition, and the Hopf bifurcation) are almost indistinguishable from each other as they occur at the very close values of $\tilde{J}_{ie}^{AMPA}$; the latter two events are better distinguished for greater values of $\tilde{J}_{ie}^{NMDA}$ (i.e. on the upper-middle branches of the diagrams 5 and 6, see below).

- After the Hopf bifurcation, alongside with destabilization of the fixed point, a stable limit cycle appears. The additional analysis demonstrated that this stable limit cycle exists only in the narrow range of $\tilde{J}_{ie}^{AMPA}$ values just below the bifurcation point. For even smaller values of $\tilde{J}_{ie}^{AMPA}$, the limit cycle disappears via homoclinic bifurcation, merging with the separatrix loop of the saddle point that corresponds to the lower-middle branch of a 1-D diagram.

- Eventually, the line 4 in the Fig. 7(a) crosses the black line AB. This corresponds to saddle-node bifurcation, analogous to the one described above for the diagrams 1 – 3. In the Fig. 7(c,d), the lower stable branch and the lower-middle unstable branch merge and disappear at this bifurcation. For smaller values of $\tilde{J}_{ie}^{AMPA}$, there are no stable fixed points in the system.

- Next, the line 4 in the Fig. 7(a) crosses the blue line that marks the border of instability of the unforced system. At this point, the diagram 4 in the Fig. 7(c,d) changes its color from blue to grey.

- After this, the line 4 in the Fig. 7(a) crosses the green line second time, and the fixed point corresponding to the upper-middle branch of the diagram 4 in the Fig. 7(c,d), transforms from unstable focus to unstable node (denoted by the dashed line without superimposed thick line).

- Finally, the line 4 in the Fig. 7(a) crosses the black line CD of saddle-node bifurcations. At this moment, the upper and the upper-middle branches of the diagram 4 in the Fig. 7(c,d) (corresponding to a saddle and an unstable node, respectively) merge and disappear. For smaller values of $\tilde{J}_{ie}^{AMPA}$, there are no fixed points in the system.

For subsequent values of $\tilde{J}_{ie}^{NMDA}$ below the cusp point B, e.g. for the values 5 and 6, the 1-D diagrams are similar to the diagram 4, with some differences. First, one can see that the rightmost parts of the lines 5 and 6 in the Fig. 7(a) lie above the green line, in the "oscillatory" region. Consequently, the fixed point that corresponds to the lower branch of the diagram 5 or 6 in the Fig. 7(c,d) for large values of $\tilde{J}_{ie}^{AMPA}$ corresponds to a stable focus, and not to a fixed node, as it was in the diagram 4. This focus transforms to a node as the line 5 or 6 in the Fig. 7(a) crosses the right part of the green line. Also, the saddle-node bifurcation that corresponds to the line DC in the Fig. 7(a) is no longer observed, because, for large $\tilde{J}_{ie}^{NMDA}$ values, the DC line goes in the physically non-sense region of negative $\tilde{J}_{ie}^{AMPA}$ values.

The important difference of the value 6 from the value 5 is that the Hopf bifurcation [crossing of the red line by the line 6 in the Fig. 7(a)] occurs in this case for a smaller value of $\tilde{J}_{ie}^{AMPA}$ than the saddle-node bifurcation [crossing of the line AB by the line 6 in the Fig. 7(a)]. This is true for all values of $\tilde{J}_{ie}^{NMDA}$ that lie between the points E and B in the Fig. 7(a). In the Fig. 7(c,d), one can see that, as we move along the diagram 6 from the right to the left, the lower two branches merge and disappear via the saddle-node bifurcation, but for somewhat smaller $\tilde{J}_{ie}^{AMPA}$ values, the system would not diverge, but jump to the upper-middle branch, as this branch is stable.

The value 7 corresponds to the cusp point B in the Figure Fig. 7(a,c,d). For $\tilde{J}_{ie}^{NMDA}$ values above this point, such as for the values 8 and 9, the saddle-node bifurcations that corresponded to the lines AB and BC in the Fig. 7(a) are no longer present. In the figure Fig. 7(c,d), one can see that the lower stable branches of the diagrams 8 and 9 do not disappear via saddle-node bifurcation as $\tilde{J}_{ie}^{AMPA}$ decreases (like it was observed for the diagrams 1 – 6). Instead of it, the lower branch continues to the left, and the fixed point that corresponds to this branch, first, transforms to a stable focus and then loses its stability via the Hopf bifurcation.

For the value 8, as for all the previous values, there is a range of $\tilde{J}_{ie}^{AMPA}$ values, for which the lower branch of the diagram 8 in the Figure Fig. 7(c,d) corresponds to a stable node. In the Fig. 7(a), the line 8 goes below the green line in this range. For the $\tilde{J}_{ie}^{NMDA}$ values above the point F in the Fig. 7(a), such as for the value 9, the lower branch of the 1-D diagram corresponds to a stable focus for all $\tilde{J}_{ie}^{AMPA}$ values to the right of the Hopf bifurcation.

**Acknowledgements:** Supported by Russian Science Foundation grant (No: 17-11-01273)